%% file: mnras_template.tex
\documentclass[fleqn,usenatbib]{mnras}
\usepackage{newtxtext,newtxmath}
\usepackage[T1]{fontenc}

\DeclareRobustCommand{\VAN}[3]{#2}
\let\VANthebibliography\thebibliography
\def\thebibliography{\DeclareRobustCommand{\VAN}[3]{##3}\VANthebibliography}

\usepackage{graphicx}	% Including figure files
\usepackage{amsmath}	% Advanced maths commands
\usepackage[flushleft]{threeparttable}
\usepackage{makecell}
\usepackage{flushend}  % To balance columns on last page
\usepackage{lscape}
\usepackage{rotating}
\usepackage{float}

\newcommand{\sm}{M$_{\odot}$}
\newcommand{\kms}{\,km\,s$^{-1}$}

\title[GASTON-GP: catalogue and variability study]{GASTON-GP: Source catalogue and millimetre variability of massive protostellar objects}

\author[Ji-Xuan Zhou et al.]{
Ji-Xuan Zhou,$^{1}$\thanks{E-mail: ZhouJ71@cardiff.ac.uk (JXZ)}
Nicolas Peretto,$^{1}$\thanks{E-mail: nicolas.peretto@astro.cf.ac.uk (NP)}
A. J. Rigby$^{2}$, 
R.~Adam$^{3}$,
P.~Ade$^{1}$,
H.~Ajeddig$^{4}$,
S.~Amarantidis$^{5}$,
\newauthor
P.~Andr\'e$^{4}$,
H.~Aussel$^{4}$,
A. Bacmann$^{12}$,
A.~Beelen$^{6}$,
A.~Beno\^it$^{7}$,
S.~Berta$^{9}$,
M.~B\'ethermin$^{8}$,
A.~Bongiovanni$^{5}$,
\newauthor
J.~Bounmy$^{10}$,
O.~Bourrion$^{10}$,
M.~Calvo$^{7}$,
A.~Catalano$^{10}$,
D.~Ch\'erouvrier$^{10}$,
M.~De~Petris$^{11}$,
F.-X.~D\'esert$^{12}$,
\newauthor
S.~Doyle$^{1}$,
E.~F.~C.~Driessen$^{9}$,
G.~Ejlali$^{13}$,
A.~Ferragamo$^{11}$,
A.~Gomez$^{14}$,
J.~Goupy$^{7}$,
C.~Hanser$^{15}$,
\newauthor
S.~Katsioli$^{16,17}$,
F.~K\'eruzor\'e$^{18}$,
C.~Kramer$^{9}$,
B.~Ladjelate$^{5}$,
G.~Lagache$^{6}$,
S.~Leclercq$^{9}$,
J.-F.~Lestrade$^{19}$,
\newauthor
J.~F.~Mac\'ias-P\'erez$^{10}$,
S.~C.~Madden$^{4}$,
A.~Maury$^{20,\ 21,\ 4}$,
F.~Mayet$^{10}$,
A.~Monfardini$^{7}$,
A.~Moyer-Anin$^{10}$,
\newauthor
M.~Mu\~noz-Echeverr\'ia$^{22}$,
I.~Myserlis$^{5}$,
Q. Nguyen-Luong$^{29,\ 4}$,
A.~Paliwal$^{23}$,
L.~Perotto$^{10}$,
G.~Pisano$^{11}$,
\newauthor
N.~Ponthieu$^{12,\ 6}$,
V.~Rev\'eret$^{4}$,
A.~Ritacco$^{10}$,
H.~Roussel$^{26}$,
F.~Ruppin$^{27}$,
M.~S\'anchez-Portal$^{5}$,
S.~Savorgnano$^{10}$,
\newauthor
K.~Schuster$^{9}$,
A.~Sievers$^{5}$,
C.~Tucker$^{1}$,
R.~Zylka$^{9}$\\
Affiliations are listed at the end of the paper
}
\date{Accepted XXX. Received YYY; in original form ZZZ}

\pubyear{2023}

\begin{document}
\label{firstpage}
\pagerange{\pageref{firstpage}--\pageref{lastpage}}
\maketitle

\input{tex/0_abstract}
% \begin{abstract}

% \end{abstract}

% % Select between one and six entries from the list of approved keywords.
% % Don't make up new ones.
% \begin{keywords}
% keyword1 -- keyword2 -- keyword3
% \end{keywords}

%%%%%%%%%%%%%%%%%%%%%%%%%%%%%%%%%%%%%%%%%%%%%%%%%%

%%%%%%%%%%%%%%%%% BODY OF PAPER %%%%%%%%%%%%%%%%%%

%\section{Introduction}
\input{tex/1_intro}

\input{tex/2_data}

\input{tex/3_method}

\input{tex/4_results}

\input{tex/5_discussion}

\input{tex/6_Summary}

\section*{Acknowledgements}
%We appreciate the helpful comments from Stefano Berta and Carsten Kramer.
JXZ would like to acknowledge the financial support from the China Scholarship Council (CSC PhD programme, No. 202207030001). NP and AJR would like to thank the STFC for financial support under the consolidated grant number ST/N000706/1 and ST/S00033X/1. The work made use of the GASTON-GP survey (PI: Nicolas Peretto).
We would like to thank the IRAM staff for their support during the observation campaigns. The NIKA2 dilution cryostat has been designed and built at the Institut N\'eel. In particular, we acknowledge the crucial contribution of the Cryogenics Group, and in particular Gregory Garde, Henri Rodenas, Jean-Paul Leggeri, Philippe Camus. This work has been partially funded by the Foundation Nanoscience Grenoble and the LabEx FOCUS ANR-11-LABX-0013. This work is supported by the French National Research Agency under the contracts "MKIDS", "NIKA" and ANR-15-CE31-0017 and in the framework of the "Investissements d’avenir” program (ANR-15-IDEX-02). This work has been supported by the GIS KIDs. This work has benefited from the support of the European Research Council Advanced Grant ORISTARS under the European Union’s Seventh Framework Programme (Grant agreement No. 291294). R. A. acknowledges support from the Programme National Cosmology et Galaxies (PNCG) of CNRS/INSU with INP and IN2P3, co-funded by CEA and CNES. R. A. was supported by the French government through the France 2030 investment plan managed by the National Research Agency (ANR), as part of the Initiative of Excellence of Université Côte d'Azur under reference number ANR-15-IDEX-01. A. Maury acknowledges support the funding from the European Research Council (ERC) under the European Union’s Horizon 2020 research and innovation programme (Grant agreement No. 101098309 - PEBBLES).
The NIKA2 data was processed using the Pointing and Imaging In Continuum (PIIC) software, developed by Robert Zylka and Stefano Berta at the Institut de Radioastronomie Millim\'{e}trique (IRAM) and distributed by IRAM via the GILDAS pages. PIIC is the extension of the MOPSIC data reduction software to the case of NIKA2 data. This publication also makes use of data products from NEOWISE, which is a project of the Jet Propulsion Laboratory/California Institute of Technology, funded by the Planetary Science Division of the National Aeronautics and Space Administration. This research used the \textsc{Python} programming language and the following software packages: \textsc{NumPy}, \textsc{SciPy}, \textsc{Astrodendro}, \textsc{BD\_wrapper}, \textsc{BTS}, \textsc{matplotlib} and \textsc{Nebuliser}.

\section*{Data Availability}
All survey products are publicly accessible on the IRAM Data Management System.  These include the deep maps, the maps of the 11 individual runs, RMS maps, and the compact source catalogues at 1.15 mm and 2.00 mm.



\bibliographystyle{mnras}
\bibliography{GASTON} % if your bibtex file is called example.bib

%\begin{thebibliography}{99}
%\end{thebibliography}

%%%%%%%%%%%%%%%%%%%%%%%%%%%%%%%%%%%%%%%%%%%%%%%%%%

%%%%%%%%%%%%%%%%% APPENDICES %%%%%%%%%%%%%%%%%%%%%
\input{tex/table115}

\appendix
% \section{Some extra material}

\input{tex/A_transfer_function}
\input{tex/B_velocity_decision}

%%%%%%%%%%%%%%%%%%%%%%%%%%%%%%%%%%%%%%%%%%%%%%%%%%

\noindent\textit{ $^{1}$School of Physics and Astronomy, Cardiff University, Queen’ s Buildings, The Parade, Cardiff CF24 3AA, UK\\
$^{2}$School of Physics and Astronomy, University of Leeds, Leeds LS2 9JT, UK\\
$^{3}$Universit\'e C\^ote d'Azur, Observatoire de la C\^ote d'Azur, CNRS, Laboratoire Lagrange, France\\
$^{4}$Universit\'e Paris Cité, Université Paris-Saclay, CEA, CNRS, AIM, F-91191 Gif-sur-Yvette, France\\
$^{5}$Institut de Radioastronomie Millim\'etrique (IRAM), Granada, Spain\\
$^{6}$Aix Marseille Univ, CNRS, CNES, LAM (Laboratoire d'Astrophysique de Marseille), Marseille, France\\
$^{7}$Institut N\'eel, CNRS, Universit\'e Grenoble Alpes, France\\
$^{8}$Université de Strasbourg, CNRS, Observatoire astronomique de Strasbourg, UMR 7550, 67000 Strasbourg, France\\
$^{9}$Institut de RadioAstronomie Millim\'etrique (IRAM), Grenoble, France\\
$^{10}$Univ. Grenoble Alpes, CNRS, Grenoble INP, LPSC-IN2P3, 53 avenue des Martyrs, 38000 Grenoble, France\\
$^{11}$Dipartimento di Fisica, Sapienza Universit\`a di Roma, Piazzale Aldo Moro 5, I-00185 Roma, Italy\\
$^{12}$Univ. Grenoble Alpes, CNRS, IPAG, 38000 Grenoble, France\\
$^{13}$Institute for Research in Fundamental Sciences (IPM), School of Astronomy, Tehran, Iran\\
$^{14}$Centro de Astrobiolog\'ia (CSIC-INTA), Torrej\'on de Ardoz, 28850 Madrid, Spain\\
$^{15}$Aix Marseille Univ, CNRS/IN2P3, CPPM, Marseille, France\\
$^{16}$National Observatory of Athens, Institute for Astronomy, Astrophysics, Space Applications and Remote Sensing, GR-15236 Athens, Greece\\
$^{17}$Department of Astrophysics, Astronomy \& Mechanics, University of Athens, GR-15784 Athens, Greece\\
$^{18}$High Energy Physics Division, Argonne National Laboratory, 9700 South Cass Avenue, Lemont, IL 60439, USA\\
$^{19}$LUX, Observatoire de Paris, PSL Research University, CNRS,
Sorbonne Université, 75014 Paris, France\\
$^{20}$Institute of Space Sciences (ICE, CSIC), Campus UAB, E-08193 Barcelona, Spain\\
$^{21}$ICREA, Pg. Lluís Companys 23, Barcelona, Spain\\
$^{22}$IRAP, CNRS, Université de Toulouse, CNES, UT3-UPS, Toulouse, France\\
$^{23}$Dipartimento di Fisica, Universit\`a di Roma ‘Tor Vergata’, Via della Ricerca Scientifica 1, I-00133 Roma, Italy\\
$^{24}$Laboratoire de Physique de l’\'Ecole Normale Sup\'erieure, ENS, PSL Research University, CNRS, Sorbonne Universit\'e, Université de Paris, 75005 Paris, France\\
$^{25}$INAF-Osservatorio Astronomico di Cagliari, Via della Scienza 5, 09047 Selargius, Italy\\
$^{26}$Institut d'Astrophysique de Paris, CNRS (UMR7095), 98 bis boulevard Arago, 75014 Paris, France\\
$^{27}$University of Lyon, UCB Lyon 1, CNRS/IN2P3, IP2I, 69622 Villeurbanne, France\\
$^{28}$University Federico II, Naples, Italy\\
$^{29}$Institute for Astronomy and Space Quantum Communications, Gia Lai, Vietnam
}

% Don't change these lines
%\bsp	% typesetting comment
\label{lastpage}
\end{document}

%% file: tex/0_abstract.tex
\begin{abstract}

\noindent The processes governing protostellar mass growth remain debated, although episodic accretion is now understood as a key feature of protostellar evolution across all masses. Luminosity bursts have been observed in both low- and high-mass protostars, but the overall statistics remain limited, especially for high-mass objects. Over the past decade, numerical simulations of high-mass core collapse have provided a theoretical framework for interpreting protostellar variability, yet additional observational constraints are required to determine the characteristics and importance of bursts. In this work, we analyse data from GASTON-GP programme, which mapped a 2.4 deg$^2$ region of the Galactic plane (centred at $l = 24^{\circ}$) at 1.15 and 2.00 mm using NIKA2 on the IRAM 30 m telescope. The survey obtained 11 epochs over four years, offering the first opportunity to study millimetre variability in a large sample of massive protostellar sources. From the combined dataset, we constructed catalogues of 2925 compact sources at 1.15 mm and 1713 at 2.00 mm. Using a dedicated relative calibration scheme, we generated millimetre light curves for ~200 high–signal-to-noise sources and identified one variable candidate. However, it is not protostellar. Consequently, we report no robust detections of variable protostellar sources in GASTON field. This is the direct consequence of observational limitations (i.e., sensitivity, resolution) combined with the lack of any > 100-fold luminosity bursts during the observations, which is consistent with estimates inferred from isolated core collapse simulations. This study highlights the need for future high-resolution, high-cadence surveys to constrain the accretion histories of massive protostars.

\end{abstract}

\begin{keywords}
surveys -- stars: formation -- stars: massive -- ISM: structure -- radio continuum: transients.
\end{keywords}

%% file: tex/1_intro.tex
%%%%%%%introduction
\section{Introduction}

%Star formation is one of the most fundamental process in astrophysics. 
Star formation determines the rate at which cold interstellar gas is converted into stars, along with the shape of their mass distribution at birth, i.e. the initial mass function (IMF) \cite[e.g.][]{Mckee2007,Bastian2010,Andre2014,Offner2014,Motte2018,Krumholz2019}. Together, those two aspects of star formation govern the evolution of galaxies across the Universe. But despite the importance of star formation, the deceptively simple question of how stars gain their mass is a matter of intensive ongoing research.

Five decades of infrared to radio observations of the sky have shown that stars form in molecular clouds, deep within their coldest ($\sim10$~K) and most compact ($\sim0.01-0.1$pc) over-densities, known as cores \citep[e.g.][]{Myers1983, Ward-Thompson1995, Motte1998, Andre2000, Bergin2007}. Protostars, identified via the observation of compact mid-infrared emission and/or high-velocity jets and outflows, appear to be deeply embedded within those cores \citep[e.g.][]{Beichman1986, Shu1987, lada1987, lada1993, Andre1993, Bontemps1996}. As a result of conservation of angular momentum, a circumstellar disc is formed that transports material to the growing protostar \citep{larson2003, Mckee2007}. In early theoretical models of star formation, it was proposed that the accretion from the core to the protostar was constant, and set by one parameter, i.e. the temperature of the gas \citep{Larson1969, shu1977}. In those models, mostly motivated by the early observation of nearby low-mass star-forming clouds, the core was taken as a fixed reservoir of mass, isolated from the larger scale molecular clouds. However, the census of nearby young stars by the Infrared Astronomical Satellite (IRAS) revealed that most of the low-mass protostars in the Taurus-Auriga molecular cloud have luminosities 10 to 50 times lower than what is predicted under steady accretion. This is known as the "luminosity problem" \citep{kenyon1990, kenyon1994}. Subsequent higher-resolution infrared surveys of nearby star-forming regions confirmed this inconsistency between the expected and observed luminosity distributions of protostars \citep{Dunham2008, Dunham2013}.

A solution to the luminosity problem is episodic accretion \citep{kenyon1990, Evans2009}. Instead of having a steady accretion process, the protostars accrete material from the disc at a low rate during most of the time, but occasionally undergo large accretion bursts. 
Multi-epoch optical and near-IR variability studies have identified young stellar objects that display long-duration outbursts (from months to 100 years) that are driven by large changes in the accretion rate (from $10^{-7}$ to up to $10^{-4}$ \sm  year$^{-1}$, see \citet{Herbig1966, Herbig1977, Herbig2007, Herbig2008, Audard2014, Fischer2023}). Statistical studies indicate that outbursts are more frequent during the younger Class0/I stages and are rare during the final stages of accretion \citep{Pena2019, Park2021, Zakri2022, Pena2024}. To get a better understanding of how important episodic accretion is to the build up of stellar masses, one thus needs to study the variability of protostellar objects at the earliest stages of their evolution. However, due to their embedded nature, variability studies of protostars are difficult. The theoretical modelling of an episodically accreting protostar embedded within a core by \citet{Johnstone2013} and \citet{MacFarlane2019} has shown that the accretion bursts are most easily detected in the mid- to far-infrared around the peak wavelength.
However, even though weaker, flux variability at (sub-)mm wavelengths is still detectable. 

To constrain the amplitude and timescale of protostellar variability at submillimetre wavelengths, the James Clerk Maxwell Telescope SCUBA-2 Transient Survey was launched \citep{Mairs2017a}. The survey conducted regular flux monitoring of eight nearby low-mass star-forming regions, including Perseus, Orion, Ophiuchus, and Serpens, at 450 and 850 $\mu$m, with a typical monthly cadence over a period of approximately six years. Over the past decade, continuous variability studies of hundreds to thousands of bright submillimetre sources have been carried out using these data \citep{Herczeg2017, Yoo2017, Mairs2017b, Johnstone2018, Mairs2019, Lee2021, Yoon2022, Johnstone2022, Mairs2024, Park2024, Sheehan2025, Chen2025}.
In particular, \citet{Lee2021} analysed the variability of $\sim300$ bright submillimetre sources and found that approximately 20\% of the 83 monitored class 0 / I protostars exhibit secular variability at 850 $\mu$m. In 2020, the observing campaign was expanded to include six intermediate- to high-mass star-forming regions ($d\leq 2$~kpc) with known accretion outbursts. Using the extended six-year 850 $\mu$m dataset, \citet{Mairs2024} identified 20 robust variable sources and 18 variable candidates, with four robust variables detected at both 850 and 450 µm, supporting an accretion-driven origin for the observed variability.

However, one can naturally expect accretion bursts towards massive protostars to be of larger amplitude and possibly more frequent. The recent findings by \citet{Rigby2024} and \citet{Zhang2024} of very complex and dynamic gas flows towards the centre of cluster-forming clumps suggest that massive protostars draw their mass from much larger mass reservoirs than the cores in which they sit, and that accretion in those systems is likely to be non-steady and proceed via successive bursts. However, massive protostars are far less numerous than low-mass ones, and they live in very crowded environments, making the characterisation of their (sub-)mm variability even more difficult. To date, bursts have been observed in only six massive protostars. The first reported accretion burst occurred in S255IR NIR3, where a $\sim$20 \sm\ protostar with a disc-like rotating structure and outflows showed a near-infrared luminosity increase. During the two-year accretion burst, its luminosity increased by a factor of 5.5 \citep{Caratti2017}. 
A more dramatic event was later reported in NGC6334-MM1. By comparing ALMA 1.3 mm observations in 2008 and 2015, \citet{Hunter2017} found a luminosity enhancement by a factor of approximately 70 in 2015 data. Masers flaring from early 2015 to 2021 indicates that the burst has persisted more than 6 years \citep{Hunter2021}. 
Further cases include: two mid-infrared bursts in M17 MIR, separated by a six-year quiescent interval \citep{Chen2021}; a far-infrared burst with methanol flares in G358 \citep{Stecklum2021}; a mid-infrared burst with methanol masers in G24.33+0.14 \citep{Hirota2022}; an 8.4-year-long burst in G323.46–0.08 with a $K_s$ band luminosity increase of 2.5 magnitudes \citep{Wolf2024}; a year-long burst showing a 68\% flux increase at 1.3 mm in I13111-6228 \citep{Yang2025}.
These observations highlight the diversity of burst durations and amplitudes for massive protostars, suggesting different mechanisms and complex accretion processes.

The physical mechanisms behind accretion bursts during star formation is still under investigation. For low-mass protostars, numerous simulations have been conducted to explore the origins of episodic accretion, which includes thermal instabilities (TI) \citep{Bell1994, Kley1999}, gravitational instability (GI) \citep{Vorobyov2005, Vorobyov2006}, the combination of GI and magnetorotational instability (MRI) \citep{Armitage2001, Zhu2009},  GI and fragmentation model \citep{Vorobyov2015} and tidal interactions \citep{Pfalzner2008}. Other factors such as disc metallicity and magnitude and direction of the rotation of the external environment can also influence the burst behaviour \citep{Vorobyov2015, Vorobyov2020}. 
%In \citet{Vorobyov2021}, they modelled accretion bursts in discs driven by three distinct mechanisms, i.e., MRI, clump infall and close stellar encounter. Different disc kinematic features have been found under different scenarios, and so observations may help to identify the triggering mechanism.}

While episodic accretion has been extensively studied in the context of low-mass protostars, simulations also began to explore its occurrence and driving mechanisms in the formation of high-mass stars. Before the first detection of the accretion burst in S255IR NIR3, radiation hydrodynamic simulations of high-mass star formation have shown variable accretion during the protostar formation process \citep{Krumholz2007, Krumholz2009, Peters2010, Kuiper2011, Klassen2016}. Regarding triggering mechanisms, a series of 3D radiation hydrodynamics simulations by \citet{Meyer2017, Meyer2019, Meyer2021} demonstrated that GI in massive protostellar discs leads to fragmentation, with infalling clumps triggering luminous accretion bursts that contribute significantly to stellar mass growth. 
In particular, \citet{Meyer2019} modelled $100\ M_\odot$ solid-body rotating cores (initial temperature 10 K), exploring different inner sink cell radii and ratios of rotational to gravitational energy ($\beta$). Based on burst statistics across models, they estimated that more than 50\% of the stellar mass is accreted during short eruptive time lasting only $\sim10^3$ years (about 1.7\% of the total simulation time).
\citet{Meyer2021} extended these studies to a broader parameter space, including core masses from 60 to $200,M_\odot$ and $\beta$ values from 0.005 to 0.33. They found that more massive and more rapidly rotating cores are more likely to undergo accretion bursts. A bimodal distribution of the burst interval, i.e., a short interval of 1 to 10 years for faint bursts and a long one of $10^3$ to $10^4$ years, were found. They may correspond to the knotty structures or giant bow shock at the top of the jets found in observation. In addition to GI, TI has been proposed as a triggering mechanism. Using 1D Shakura-Sunyaev viscous disc model, \citet{Elbakyan2024} simulated the TI outburst in high-mass YSO accretion disc. Stellar mass (5, 10, 20 \sm), accretion rate onto the disc ($3\times10^{-5}$, $10^{-4}$, $3\times10^{-4}$) and the disc viscosity ($\alpha_{\rm cold} = 0.01$, $\alpha_{\rm hot} = 0.1$) varied in different models. While some bursts matched observed durations and peak rates, the model struggled to reproduce short-duration events, suggesting TI alone may not be sufficient. 
Earlier, \citet{Elbakyan2021} employed a 1D time-dependent viscous disc evolution code to compare TI, MRI, and planet disruption. Their results indicated that neither TI nor MRI could reproduce observed burst properties, while tidal disruption events work better in certain cases.
In addition to disc accretion, recent observations suggested that the large-scale infall from outside the core through streamers can drive the accretion variability. \citet{Pineda2020} found a streamer outside the core (> 10000 AU), whose infall rate is comparable to the current accretion rate estimated from the luminosity. Their finding suggests that the previous burst can be caused by accretion from outside the envelope. 
Similar streamer structures were also observed towards the massive young stellar objects G11.92 MM2 \citep{Sanhueza2025} and GGD27-MM1 \citep{2023ApJ...956...82F}, and 7 low-mass Class 0 or Class I protostars in NGC 1333 SE fibers \citep{Valdivia-Mena2024}. These observations indicate the potential role of external accretion flows, in addition to accretion through discs, in driving episodic accretion events.

To make further progress in our understanding of protostellar mass growth, large and sensitive unbiased (sub-)mm surveys are necessary. Such surveys are now feasible due to the high sensitivity and mapping speed of large (sub-)mm cameras on telescopes such as the JCMT and the IRAM 30m.
In that respect, the Galactic Star Formation with NIKA2 (GASTON) large programme on the IRAM 30m  provides the ideal dataset \citep{Rigby2021a}. About a third of the GASTON time has been dedicated to the observation of a $\sim$ 2.4-deg$^2$ slice of the Galactic plane across 11 epochs between 2017 and 2021, giving us the opportunity to study protostellar variability for thousands of sources. This Galactic plane dataset is referred to as GASTON-GP in the remainder of the paper. 
In Sec. \ref{data}, we describe the GASTON-GP data. The methods we used for source extraction and calibration are presented in Sec. \ref{method}, while the properties of GASTON-GP compact sources are presented in Sec. \ref{property study}. Section \ref{variability study} explains how we study the variability. Finally, a discussion on number statistics of variable candidates and the limitations of GASTON-GP observations are presented in Sec. \ref{discussion}.

%% file: tex/2_data.tex
%%%%%%GASTON data intro
%%%%%%1-introduction about GASTON data, NIKA
%%%%%%2-introduction about the 11 data runs
%%%%%%3-a table about each data run? a picture about deep maps at both wavelengths

\section{The GASTON-GP data} \label{data}

In this paper we present the complete data set from the Galactic plane (GP) survey of the Galactic Star Formation with NIKA2 (GASTON) project (PI: N. Peretto), a 200-hour guaranteed-time large programme of observations carried out at the IRAM 30-m telescope using the New IRAM KID Array 2 \citep[NIKA2;][]{Calvo2016, Bourrion2016, Adam2018, Perotto2020} instrument.  NIKA2 is a dual-band millimetre continuum camera which can image at 260 GHz (1.15\,mm) and 150 GHz (2.00\,mm) simultaneously. It has a field of view (FoV) of 6.5 arcmin and an angular resolution of 11.6$''$ at 260 GHz, and 18$''$ at 150 GHz \citep{Adam2018, Perotto2020, piic2024}. Its kinetic inductance detectors (KIDs) provide excellent sensitivity, stability, and dynamic range \citep{Monfardini2010, Catalano2014, Adam2014}, suitable for a wide range of projects, from nearby star formation to galaxy clusters. The 200 hours of GASTON have been split into three sub-projects, each with a specific focus: massive star formation in the Galactic plane (GASTON-GP); brown dwarf formation in nearby star-forming clouds; and dust properties in nearby pre-stellar cores. The data we present here are those from GASTON-GP, which accumulated to a total of 76.3 hours of telescope time. 

\subsection{Observations}

The GASTON-GP field is centred on the $\ell=24\degr$ region of the Galactic mid-plane, which was selected for its high density of both Hi-GAL far-infrared clumps \citep{Molinari2016} and infrared dark clouds \citep{Peretto2009}. The field was observed using on-the-fly mapping with a series of scans centred on three positions at $\ell=23\fdg3$, $23\fdg9$, and $24\fdg5$, with $b=0\fdg05$. For each of the field centres, pairs of scans were obtained by scanning across the Galactic plane with a varying position angle with respect to its normal. The position angles were selected as a compromise between minimizing striping in the noise fields of the data, which is greatest where the angle between scan pairs is smaller, whilst also maximizing the amount of extended emission that could be recovered since scanning along the plane presents problems for data reduction due to the lack of emission-free baselines.

\begin{figure}
    \centering
    \includegraphics[width=\linewidth]{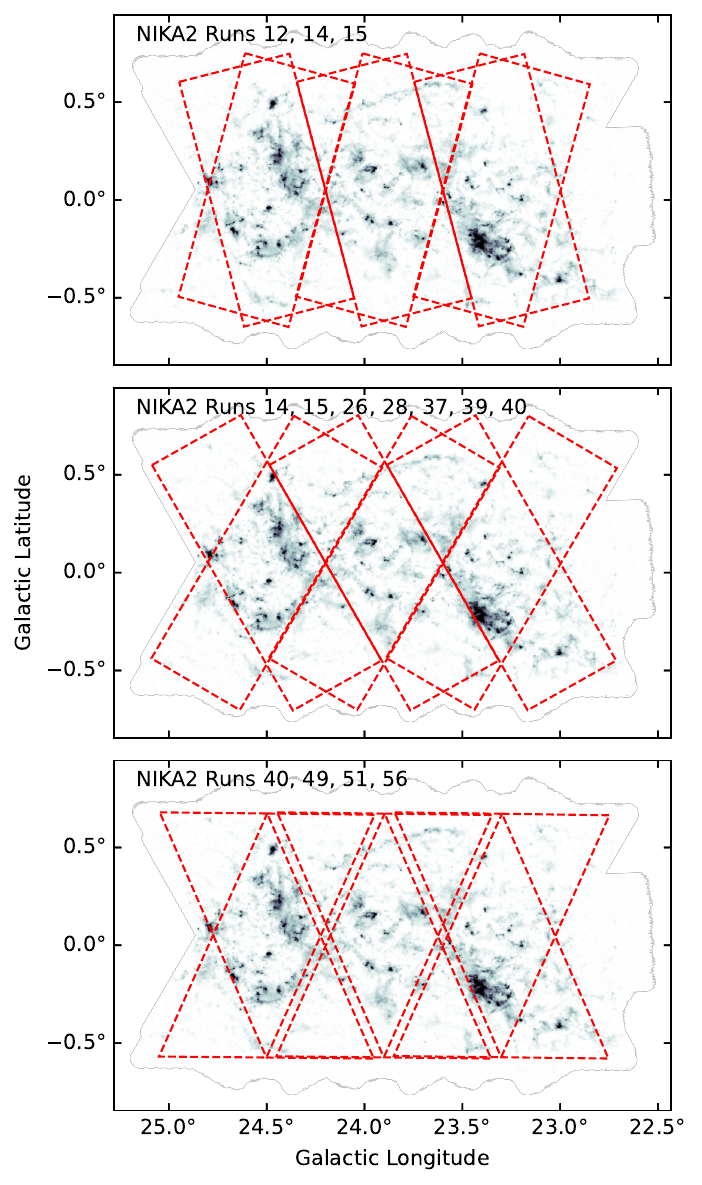}
    \caption{Scanning strategies used across the various GASTON-GP observing campaigns, overlaid on the 1.15 mm deep SNR map. The campaigns where each scan type was used are listed at the top of the corresponding panels.}
    \label{fig:scanningstrategy}
\end{figure}

The earliest scans were rectangular, and used a position angle of $\pm15$ degrees with respect to the GP normal, similar to the observations of ATLASGAL \citep{Schuller2009}, while later rectangular scans adopted a position of angle of $\pm30\deg$. The most recent scans took advantage of the \verb|tiltAngle| parameter to adopt a parallelogram observing mode with NIKA2, maintaining the $\pm30$ degrees position angles, but covering the survey area more efficiently than the earlier rectangular scans. The extent of the scans was set to cover a range in latitude of approximately $-0\fdg6$ to $0\fdg7$. Scanning speeds of 60 arcsec\,s$^{-1}$ were used in the earlier scans, but increased to 80 arcsec\,s$^{-1}$ for the later scans because faster scanning allows for better reconstruction of extended emission. We note that several scans from Run 14 were observed with an erroneous position angle due to a failed attempt to conduct the scans using the Galactic coordinate system. Here, and in the rest of this paper, "run" refers to one distinct IRAM 30m observing block, which, in the case of GASTON-GP, were 1 to 15 days long. The various scanning strategies used are illustrated in Fig. \ref{fig:scanningstrategy}, excluding the failed attempt at observing in Galactic coordinates.

The field was observed during 11 observing runs between October 2017 and November 2021 (a more detailed time line is shown in Tab. \ref{run-infor}, and these runs covered the area to differing extents. In total, the $\ell=23\fdg3$ and $24\fdg5$ positions were covered by 46 pairs of scans, while the $\ell=23\fdg9$ position was covered by 47 pairs of scans. When combined in a mosaic, the scans cover a total area of $\sim$2.4-deg$^2$.

As with all NIKA2 pool observations, pointing and focus corrections were made at regular intervals (typically hourly and 2--3 hourly, respectively). Pointing errors are conservatively estimated to have an rms of $<3$ arcsec \citep{Perotto2020}.

\subsection{Data Reduction} \label{data reduction}

% NB: Script used was mapTP_GASTON-GP.piic in
% /home/reblodata/spxar1/Projects/GASTON/DR/PIIC/red
% Deep maps in N2R12-56/
% Maps for individual runs are in N2R12/, N2R14/, etc, and are compiled in variability_Jixuan/

The GASTON-GP data were reduced using 
%version \verb|jun2025a|\footnote{\url{https://www.iram.fr/~gildas/dist/piic-exe-jun25a.tar.xz}} {\bf (NP: this is not true anymore!)check with Andy} of 
the Pointing \& Imaging in Continuum ({\sc piic}) software\footnote{\url{https://www.iram.fr/~gildas/dist/piic.pdf}} \citep{Zylka2013, piic2024} which transforms the time-series data from the NIKA2 arrays into calibrated continuum images. The input data contain records for the intensity of signal for each KID as a function of time which, when combined with knowledge of where the telescope was pointing at each instant, allows the reconstruction of a map of the source. However, while the timelines for each KID may contain astronomical signal, this is in addition to signal from the sky, as well as instrumental effects and noise from the combined telescope-and-detector system. {\sc piic} works to remove the effects of low-frequency sky noise and instrumental instabilities in order to preserve as much of the astronomical signal as possible. Many of the non-astronomical signals are correlated between the KIDs at any given instant, and the best-correlating KIDs are used to calculate the correlated noise which is then removed from all of the individual timelines. During this process, any KIDs observing an astronomical source at any given instant need to be identified, and their records removed from the analysis chain to avoid contaminating the noise model. Once a noise model is established and subtracted from the timelines, in principle only astronomical signal remains. As with all ground-based submillimetre and millimetre continuum imaging, this process necessarily results in the removal of extended astronomical signals; for NIKA2, emission on spatial scales larger than $\sim$6.5 arcminutes (the instantaneous field of view), is impossible to separate from noise, and is thus removed. We estimate that {\sc piic} is able to recover around 85 per cent of the flux on angular scales of 6.5 arcminutes, and describe our estimation of the transfer function (i.e. the level of flux recovery as a function of spatial scale) in Appendix \ref{app:transferfunction}.

% NB: Transfer Function. AJR presented a preliminary analysis to the NIKA2 telecon
% on 19/4/23. https://wiki.iram.fr/wiki/nika2/index.php?title=April_19th,_2023

Data reduction with {\sc piic} is an iterative process, in which the first step (the \emph{0th iteration}) is to perform all processing of the full data set with no a-priori definition of the astronomical source. On subsequent iterations, {\sc piic} repeats the procedure by identifying the region of the map above a user-defined SNR threshold, and removing those records from the time-series data in order to improve the measurement of the sky and instrumental noise. In this way, each subsequent iteration is able to recover increasingly faint astronomical signals. 

For the most part, the data reduction was carried out using PIIC version \verb|sep21a|\footnote{\url{https://www.iram.fr/~gildas/dist/archive/piic/}}, for which we adapted the template  {\sc piic} script for use with a crowded Galactic plane field with bright emission by setting \verb|weakSou| to \verb|no|, and altering the following key parameters:
\begin{enumerate}
    \item We set \verb|snrLevel|, which controls the threshold used to define the source model, to 4.
    \item We set \verb|nIter|, which gives the number of {\sc piic} iterations to perform, to 100.
    \item We set \verb|blOrderOrig|, the order of the polynomial fit to the baseline of each scan, to 7.
\end{enumerate}
These settings were chosen after systematic testing of the parameter space, and provide a compromise between the wish to recover as much extended emission as possible whilst avoiding generating spurious extended signals. As a second stage, we used the output model of these 100 iterations as the input model (using the \verb|sbSource| option) of a further single iteration using version \verb|jun25a| of PIIC in order to incorporate the latest improvements and updates.

The above settings were used to produce the `deep' maps -- the deepest 1.15 and 2.00 mm maps which use all suitable scans from all 11 observing runs. We also produced a map for each individual run using version \verb|jun25a| of PIIC, for which we used the model outputs from the deep maps as the source model -- again using the \verb|sbSource| option. By doing this, the individual run maps are able to capitalise on the extended source recovery already determined in the deep map processing, despite the much lower integration times of the individual runs.

% which uses the same settings with the exception of \verb|nIter| which was set to 30 \textcolor{red}{[need to check with Andy]}. This latter choice was made because the integration times for the individual runs was much lower than for the deep maps, and so the recovery of faint and extended emission converged in fewer iterations.

\begin{figure*}
\includegraphics[scale=1.05]{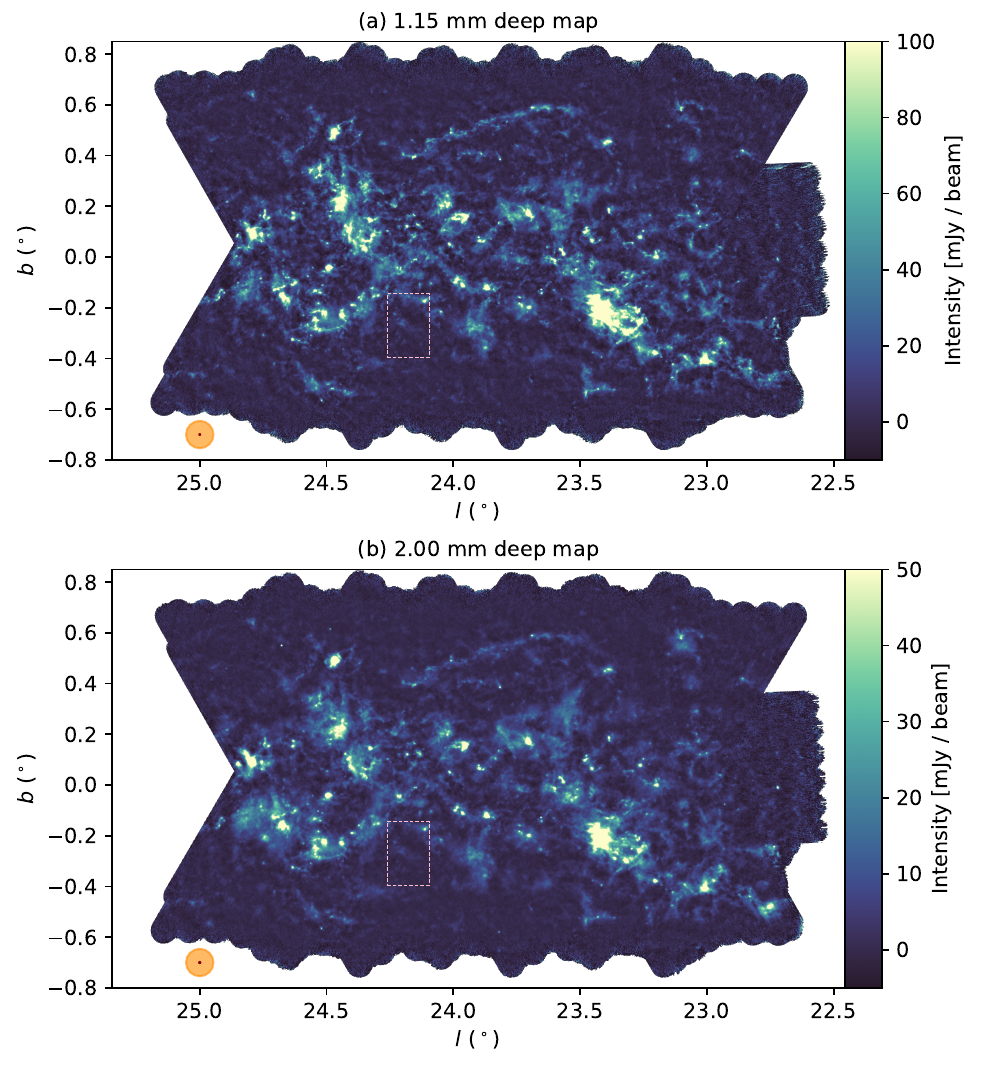}
\caption{ Panel (a): 1.15 mm deep map obtained by combining all 11 GASTON-GP runs at 1.15 mm. The orange and red circle in the left bottom marks the FoV(6.5' diameter) and beam size (11.6$''$) at 1.15mm, respectively. The pink polygon shows the area used for the calculation of the rms noise presented in Tab. \ref{run-infor}.  Panel (b): 2.00 mm deep map obtained by combining all 11 runs at 2.00 mm. The orange and red circle in the left bottom marks the FoV(6.5$'$ diameter) and beam size (18$''$) at 2.00 mm, respectively. The pink polygon is the same as that in 1.15 mm map,  where the rms noise in Tab. \ref{run-infor} is estimated.}
\label{deepmaps}
\end{figure*}

\begin{figure*}
\includegraphics[scale=0.9,angle=0]{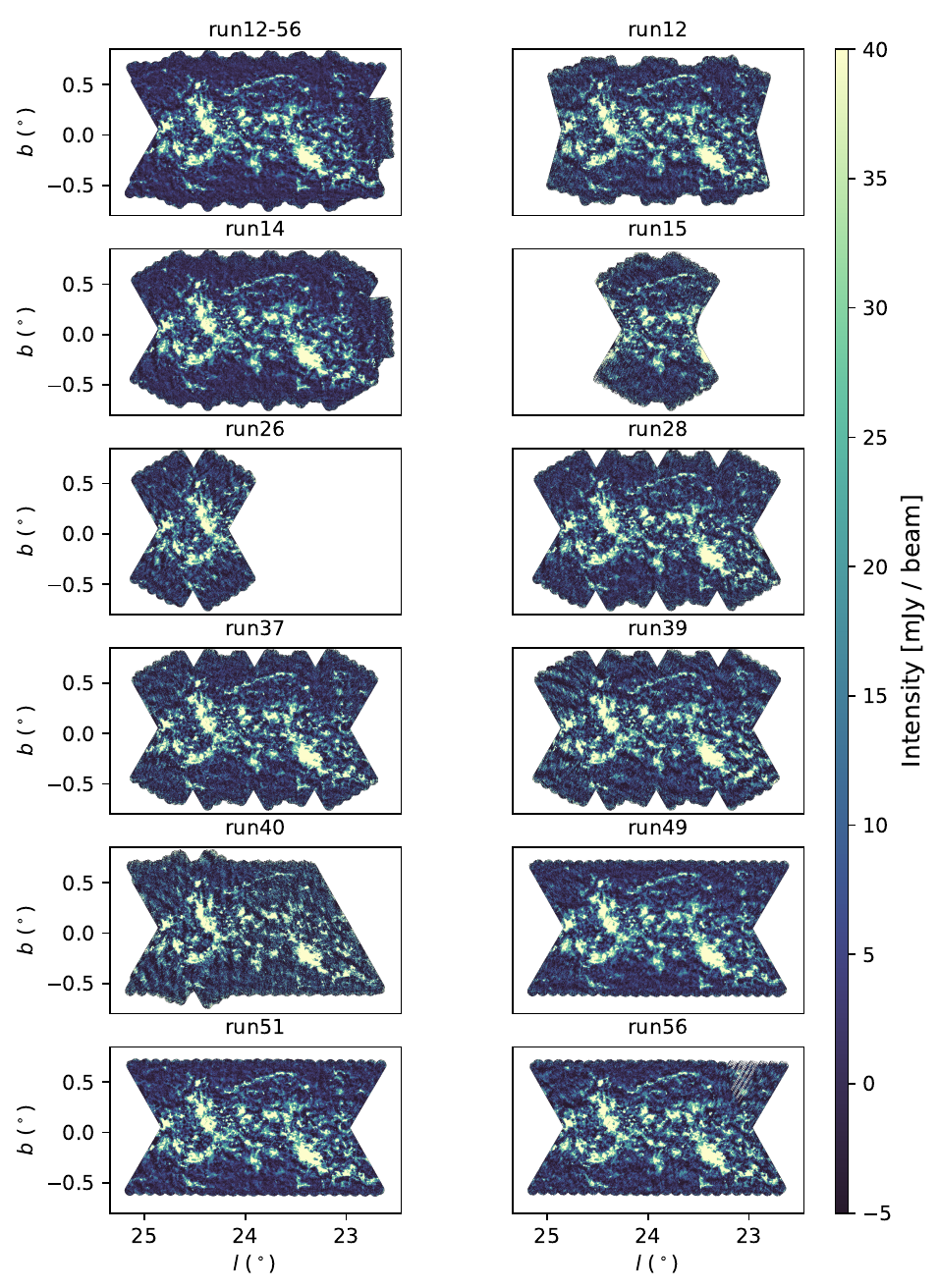}
\caption{ Deep map (top-left panel), alongside the maps obtained for each individual run at 1.15 mm. }
\label{115_11_examples}
\end{figure*}

The resulting deep maps in the two wavebands are shown in Fig. \ref{deepmaps}, and we show the 1.15\,mm maps for the individual runs in Fig. \ref{115_11_examples}. Alongside maps of the astronomical signal, {\sc piic} generates a root-mean-square (RMS) map, showing the noise level across the field. To compare the noise levels between different data runs, we selected a large region covered in all runs and computed the median RMS value within that area for each map. The resulting values are listed in Tab.~\ref{run-infor}, and the selected region is outlined as a pink polygon in Fig.~\ref{deepmaps}.

% The rms for each map was measured by {\sc piic} automatically. {\sc piic} searches for a square area in the map with an edge-size of between 5 and 8 times the beam FWHM with the lowest RMS, and uses this `noise polygon' to measure a representative noise value. We decided to apply the polygon defined in of the each deep maps to calculate the rms value in each individual run. We list the noise values for all maps in Tab. \ref{run-infor}. 
% We note that the noise polygon varies in position from map to map and between the two wavebands, since each individual run map and wavebands have different noise characteristics. 
% We further note that the reported rms values represent the sensitivity only in the deepest parts of the maps. The noise polygons used for all the maps at 1.15 and 2.00 mm
% the deep maps' noise estimates
% are illustrated in Fig. \ref{deepmaps}. 

Uranus and Neptune were used as primary flux calibrators. The resulting point-source absolute calibration uncertainty is about 5\%. To estimate the absolute calibration uncertainty on the GASTON-GP data themselves, we derived the mean and standard deviation of the brightest 36 GASTON-GP sources (see Sec. \ref{flux calibration}) across all 11 runs, and determined an absolute flux calibration uncertainty of 14\% and 7\% at 1.15 mm and 2.00 mm, respectively. Those larger uncertainties are mostly driven by the presence of significant large-scale emission and systematics in the data quality of some runs.

\begin{table} 
\caption{Observation time and rms noise across different runs. The rms values shown represent the median rms within the polygon region of each unfiltered map. The locations of the polygons are marked by pink boxes on the 1.15 and 2.00 mm maps in Fig. \ref{deepmaps}.}
\label{run-infor}
\begin{tabular}{cccc}
\hline
data run id & observation time & rms$_{\rm 1.15mm}$ & rms$_{\rm 2.00 mm}$\\
            &                  &         (mJy / beam)        &   (mJy / beam)    \\\hline
Run 12      &       27/10/2017-30/10/2017           &          7.44       &     2.07         \\
Run 14      &      17/01/2018-23/01/2018           &          4.28        &     1.26      \\
Run 15      &      17/02/2018-18/02/2018          &          11.93      &    2.87       \\
Run 26      &     18/01/2019-22/01/2019            &          9.44      &     2.79       \\
Run 28      &      13/02/2019-17/02/2019          &          7.36      &      2.06     \\
Run 37      &    08/11/2019            &          8.64      &       2.34    \\
Run 39      &    16/01/2020-17/01/2020           &          8.43      &       2.37  \\
Run 40      &     30/01/2020-04/02/2020           &          14.13      &      4.04    \\
Run 49      &     13/01/2021-17/01/2021             &          4.93      &       1.75  \\
Run 51      &     16/03/2021-22/03/2021            &          3.39       &       1.14   \\
Run 56      &     07/11/2021-26/11/2021           &          5.46       &       1.74   \\
Run 12-56  &  -    &  1.66  &     0.54        \\\hline    
\end{tabular}
\end{table}

% \begin{figure*}
% \includegraphics[scale=0.7]{pics/deepmap.pdf}
% \caption{{\bf panel (a): Deep map at 1.15 mm.} This is the combined deep map using the 4-year data of GASTON-GP at 1.15 mm. The blue and red circle in the left bottom marks the FoV(6.5' diameter) and beam size(11.1") at 1.15mm. The red square shows the polygon area we use to study the rms value at 1.15 mm. {\bf panel (b): Deep deep map at 2.00 mm.}  This is the combined deep map using the 4-year data of GASTON-GP at 2.00 mm. The blue and red circle in the left bottom marks the FoV(6.5' diameter) and beam size(17.6") at 2mm. The red square shows the polygon area we use to study the rms value at 2.00 mm. {\bf panel (c):  Filtered map at 1.15 mm.} This is the filtered deep map at 1.15mm after using a 60 arcsec filter in Nebuliser. {\bf panel (d):  Filtered map at 2.00 mm.} {\bf (NP: Using the same filetering as for 1.15mm?)}}
% \label{deep_example}
% \end{figure*}

\begin{figure*}
\includegraphics[scale=0.9]{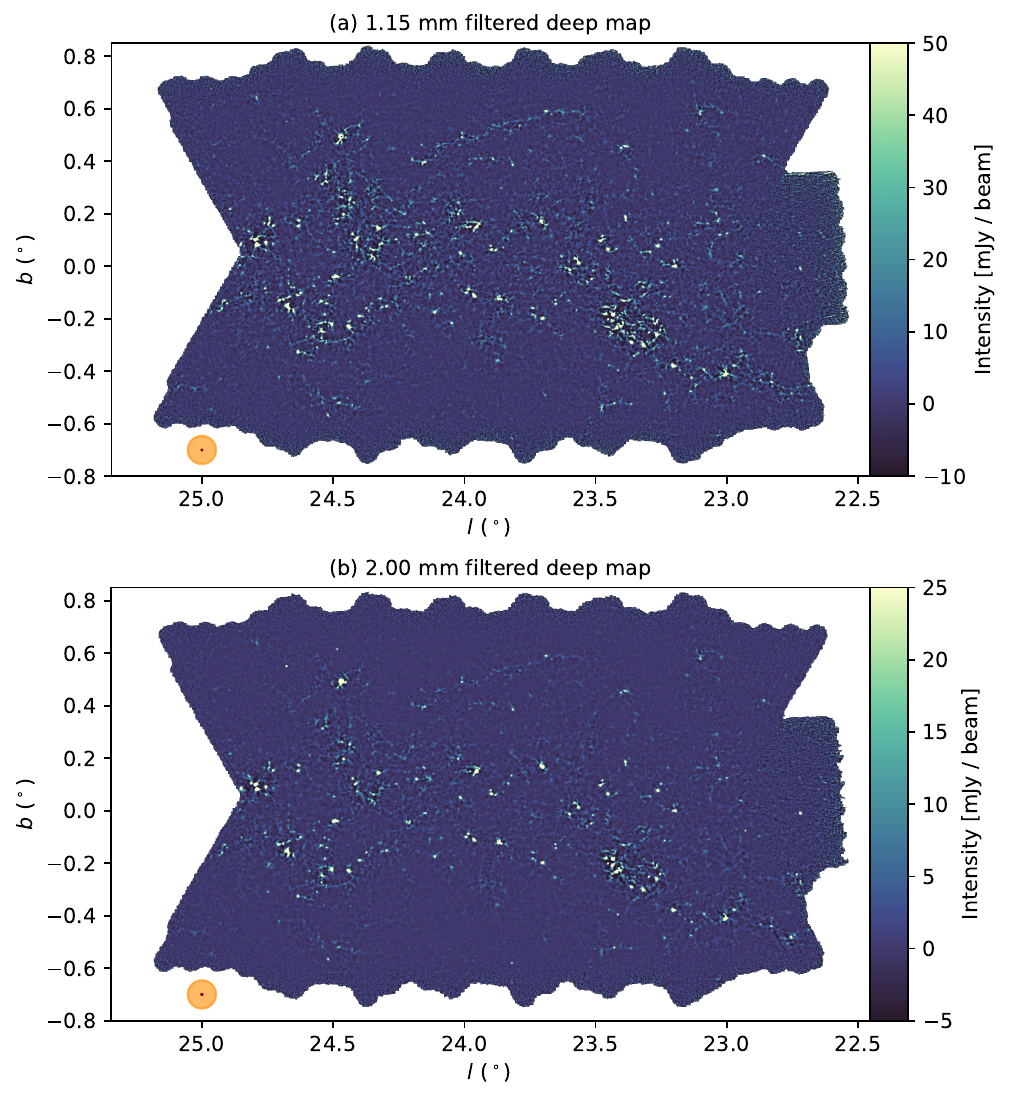}
\caption{{ Panel (a):} 1.15 mm map obtained after applying a 60-arcsec spatial filter with Nebuliser. {Panel (b):} 2.00 mm map obtained after applying a 60-arcsec spatial filter with Nebuliser. The orange and red circle in the left bottom in both panels refers to the FoV and beam size, respectively.}
\label{filtered_deepmaps}
\end{figure*}

%% file: tex/3_method.tex
\section{Compact source identification} \label{method}

\subsection{Extended emission removal with Nebuliser} \label{filter}
Given the 5.5~kpc average distance of GASTON-GP sources \citep{Rigby2021a}, our observations can only resolve structures larger than $\sim0.3$~pc and $\sim0.5$~pc at 1.15 mm and 2.00 mm, respectively. As a result, GASTON-GP compact sources are typically the size of parsec-scale clumps \citep{Williams2000}, i.e. the progenitors of individual star-clusters.  We thus gear our source identification towards the detection of such compact sources. 
%Previous studies have shown that single star formation occurs within dense cores on sub-pc scales in molecular clouds \citep{Myers1983, Williams2000, andre2014, Motte2022}. Based on the size of these cores, we focus on the very compact emissions in the data maps to trace star formation.

We first used the tool \texttt{Nebuliser}\footnote{\url{http://casu.ast.cam.ac.uk/surveys-projects/software-release/background-filtering}} to remove the extended emission in the maps. \texttt{Nebuliser} is a background-removal package that models and subtracts extended emission based on a given scale. Unlike other background estimation methods that use grid-based averaging and smoothing, \texttt{Nebuliser} employs a series of interactive sliding median and mean filters to model spatially varying backgrounds, making it more effective in complex regions.

We applied \texttt{Nebuliser} to the two deep maps and individual run maps at 1.15 and 2.00 mm to filter out extended emission. The filter size was set to 60 arcsec ($\sim$ 1.6 pc at a distance of 5.5 kpc), corresponding to 20 pixels at 1.15 mm and 15 pixels at 2.00 mm. The filtered maps were then used for both source extraction and flux density measurements. Fig. \ref{deepmaps} presents the original deep maps at 1.15 mm and 2.00 mm, and the corresponding filtered deep maps are shown in Fig. \ref{filtered_deepmaps}. With extended emission removed, we can identify the most compact sources in the maps.

To take into account the non-uniform nature of the noise within each map, we performed the compact source identification on the signal-to-noise ratio (SNR) maps, built from the ratio of the filtered maps to the RMS maps produced by {\sc piic}. To evaluate whether the filtering process affects the noise properties, we compared the RMS values measured from an emission-free region of the unfiltered and filtered deep maps. The RMS values show a difference of $\sim 10\%$, thus validating our approach of using the {\sc piic} RMS map.
% and, therefore, needs to be characterised. As part of the data reduction process, an rms map of the unfiltered data is generated.
%This noise map is derived from the on-sky integration time at each pixel, and do not take into account the additional noise from the map reconstruction process {\bf (NP: is that correct? Andy should be able to tell us.)}. 
% To derive the noise map of the filtered data, we determined an appropriate factor by which to rescale the unfiltered map. 

% Within the deep map, PIIC identified, in each band, a region in the deep map, defined by a polygon with the lowest rms value (see orange square symbol in Panels (a) and (b) of Fig.~\ref{deepmaps}). We calculated the rms values within the same polygon-defined regions in both the unfiltered and filtered data from each run: ${\rm rms_{origin}}$ and ${\rm rms_{filt}}$. The rms map for the filtered data was derived by rescaling the origin rms map by the factor $\rm{\frac{rms_{filt}}{rms_{origin}}}$. 
Then, the signal-to-noise ratio (SNR) map of the filtered data was derived by taking the ratio of the filtered map over the corresponding rms map. The SNR maps for the 1.15 and 2.00 mm filtered deep maps, referred to as deep SNR maps in the following sections, are used for source extraction.

\subsection{Dendrogram source extraction}
To extract the compact sources in the data, we used the dendrogram algorithm from the {\sc Astrodendro} Python package in both the 1.15 and 2.00 mm deep SNR maps. It extracts hierarchical structures from the image based on the data iso-contours \citep{2008ApJ...679.1338R}. There are three main input parameters for {\sc Astrodendro}: {\verb|min_value|}, {\verb|min_delta|}, and {\verb|min_pixel|}. {\verb|min_value|} sets the lowest data threshold to search for structures, {\verb|min_delta|} refers to the data difference between structures and substructures and {\verb|min_pixel|} describes the smallest size for a structure. The identified structures are divided into trunks, branches and leaves. Trunks are the outer, lower-density structures that contain all the substructures, branches are those inside trunks and contain leaves, while leaves are the smallest structures that do not contain any other sub-structures within it.

Instead of extracting structures in the filtered deep map directly, we chose to use the deep SNR map due to the changing rms noise over the observation field. The {\verb|min_value|} and {\verb|min_delta|} were set to 3, which allows us to find the significant emission in the map. The {\verb|min_pixel|} was set to half of the beam area. It is calculated by: ${\verb|min_pixel|} =  \pi \left(\frac{\theta_{\rm{beam}}^2}{8{\rm ln}2}\right)$, where $\theta_{\rm{beam}}$ is the beam FWHM in units of pixels. The {\verb|min_pixel|} parameter is 9 pixels at 1.15 mm and 12 pixels at 2.00 mm.

\begin{figure*}
\includegraphics[scale=0.99]{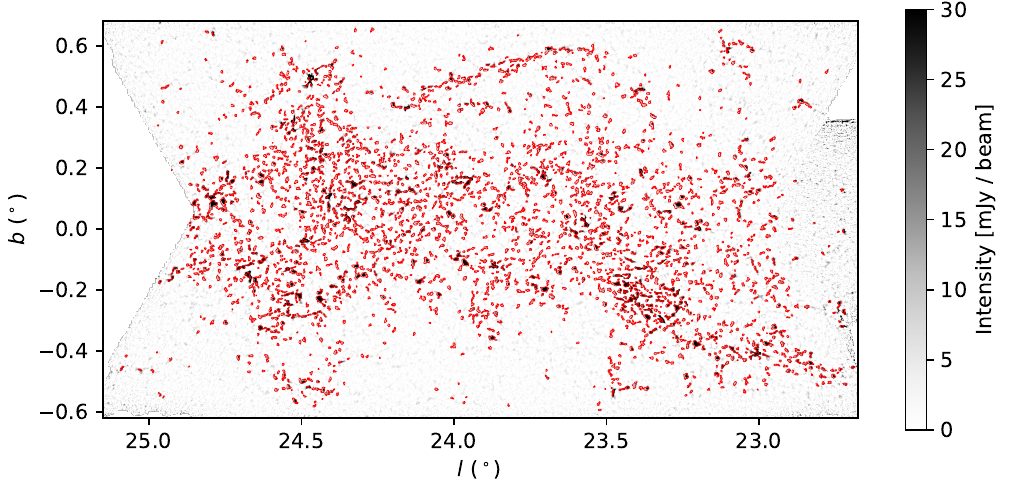}
\caption{Leaves found by {\sc Astrodendro} at 1.15 mm. The background is the 1.15 mm filtered map. Red contours mark the leaf (compact sources) footprints identified in the filtered SNR map at 1.15 mm.}
\label{leaf}
\end{figure*}

Using these parameters, a total of 3889 structures were identified in the 1.15 mm deep SNR map, of which 2925 are leaves. From the 2.00 mm deep SNR map, we extracted 2105 structures, including 1713 leaves. We refer to leaves as "compact sources" hereafter. We cross-matched the compact sources in both wavebands by checking whether the peak pixel of a 1.15 mm compact source falls onto the footprint of a 2.00 mm leaf, 1527 compact sources were detected at both wavelengths.  Fig. \ref{leaf} shows the identified leaves overlaid on the 1.15 mm filtered deep map. We present the catalogue of 1.15 mm compact sources in Tab. \ref{115table}. A similar 2mm compact source catalogue has been compiled, both being available online.

\subsection{Compact source positions and intensities}
{\sc Astrodendro} provides the footprints of all identified compact sources as pixel masks. Within the mask of each source, we located the pixel with the highest intensity, which was taken as the peak position of source $i$. Using this position, we measured the peak SNR on the filtered SNR map $({\rm SNR}_{i})$, as well as the Galactic coordinates $(l_{i},\ b_{i})$, peak intensity $(I_{i})$ on the filtered deep map, and rms noise level $({\rm RMS}_{i})$ from the RMS map. The integrated flux density $(S_{i})$ of each source was then obtained by integrating the intensities over its footprint on the filtered map.
For each observing run, we used the corresponding filtered map and re-identified the peak position within one pixel of that in the deep map, to account for a possible $\sim$3-arcsec pointing offset. Using this updated position, we recorded the signal-to-noise ratio (SNR$_{i,j}$), rms noise (RMS$_{i,j}$), and peak intensity ($I_{i,j}$) for each source at run $j$.
A summary of these quantities is provided in Table~\ref{115table}. In the following sections, we focus on the GASTON-GP compact sources. 

%and the full leaf catalogues at both wavelengths will be available online.
%with maximum flux density $(I_{i})$ within the footprints of source $i$ in the filtered deep map. For the different data runs, taking the pointing error of about 3 to 4 arcsec (about one pixel) into consideration, we allow an one-pixel offset to measure the peak flux density for each source. We use the peak location $(x_{i,\ j},\ y_{i,\ j})$ with maximum flux density $(I_{i,\ j})$ among the region $[x_i-1:y_i-1, x_i+1:y_i+1]$ as the peak location for source $i$ at run $j$. The peak flux density $(I_{i,\ j})$, SNR $({\rm SNR}_{i,\ j})$ and filtered rms $({\rm rms}_{i,\ j})$ value at these peak location are included in the catalogue. 

%Some of the structure information of the leaf structures at 1.15 mm are presented in Tab. \ref{1.15mm-cata}, and the full leaf catalogues at both wavelengths will be available online.

\section{Physical properties of compact sources} \label{property study}

Two key properties of any star-forming clump are its size and mass. In this section we describe the method we used to derive the velocities, kinematics distances, and dust temperatures for the GASTON-GP compact sources, with which we calculate their sizes and masses.  

\subsection{Radial velocity}
The first step towards deriving the physical properties of the GASTON-GP compact sources is to measure their systemic velocities, as for most Galactic plane sources, kinematic distances are the only distances one has access to.  
%Radial velocity is an important property for the dense structures in the molecular clouds. It can help to infer more kinematic information and physical properties. 
Considering the dense nature of the leaves, we derived their velocities using a set of tracers, including different isotopologues of CO, and NH$_{3}$. In most cases, multiple velocity components are present along the line of sight, and assigning the correct velocity to the source is not always straightforward. Here, we followed a similar strategy as presented in \cite{Rigby2021a} and more details can be found in Appendix \ref{velocity_decision}. Note that the spectral line data that are used here have angular resolutions that are worse by a factor  $<3$ compared to those of the GASTON-GP observations. Although this means that more structures may enter the beams of the line data, it is safe to assume that, in the vast majority of cases, the emission line is still dominated by the central GASTON-GP sources.

Leaves are the densest non-fragmented structures within the GASTON-GP dataset at the resolution of the observation. As a result, we can reasonably expect that those are most accurately identified in the densest gas tracers. We thus started with the NH$_3$ (1, 1) data from the Radio Ammonia Mid-Plane Survey \citep[RAMPS-][]{Hogge2018}. Observed with the 100 m diameter Robert C. Byrd GBT \citep{Prestage2009}, the RAMPS data have an angular and a velocity resolution of 32 arcseconds and 0.018 km s$^{-1}$, respectively. As a well-known tracer of dense gas, ammonia has a critical density of $\sim\ 3\times10^{3}$ cm$^{-3}$ \citep{Evans1999} and is therefore well-matched to the density of GASTON-GP clumps. In the situation where the ammonia line at the GASTON-GP peak location is strong enough, we used the velocity derived from the line fitting provided by \citep{Hogge2018}. If ammonia emission is present but weak we use their moment 1 velocity. Compact sources with ammonia velocities are flagged as 0.
% \textcolor{red}{There is indeed a mismatch in the angular resolution of two datasets. However, this difference does not significantly affect the derived velocities.}

For sources with no RAMPS detection we used data from the Nobeyama 45m FOREST Unbiased Galactic plane Imaging survey (FUGIN, \citet{Umemoto2017}).
FUGIN has observed both the 1st and 3rd quadrants of the Galaxy in C$^{18}$O(1$-$0), $^{13}$CO(1$-$0) and $^{12}$CO(1$-$0). With a beam size of 14$''$ for $^{12}$CO, 15$''$ for $^{13}$CO and C$^{18}$O, and a velocity resolution of 1.3 \kms, the FUGIN data are well adapted to the velocity characterization of GASTON-GP sources. 
We reprojected the three CO cubes onto the GASTON-GP pixel grid and calculated a weighted mean spectrum, weighted by the 1.15 mm intensity within the footprint for each compact source, which was then used to fit the velocity components using the automated multi-velocity Gaussian component fitting tool BTS \citep{Clarke2018}. Considering that among the CO isotopologues, C$^{18}$O traces the highest-density region and $^{12}$CO traces the most diffuse region, we selected velocities in the order of C$^{18}$O, $^{13}$CO, and $^{12}$CO. Based on the features of the spectra, e.g., single or multiple components, velocity difference between components and velocity dispersion, we assigned different velocities to the sources and flagged them from 1 to 7. A detailed description of the velocity decision is provided in the Appendix \ref{velocity_decision}.

\begin{figure*}
\includegraphics[scale=0.68]{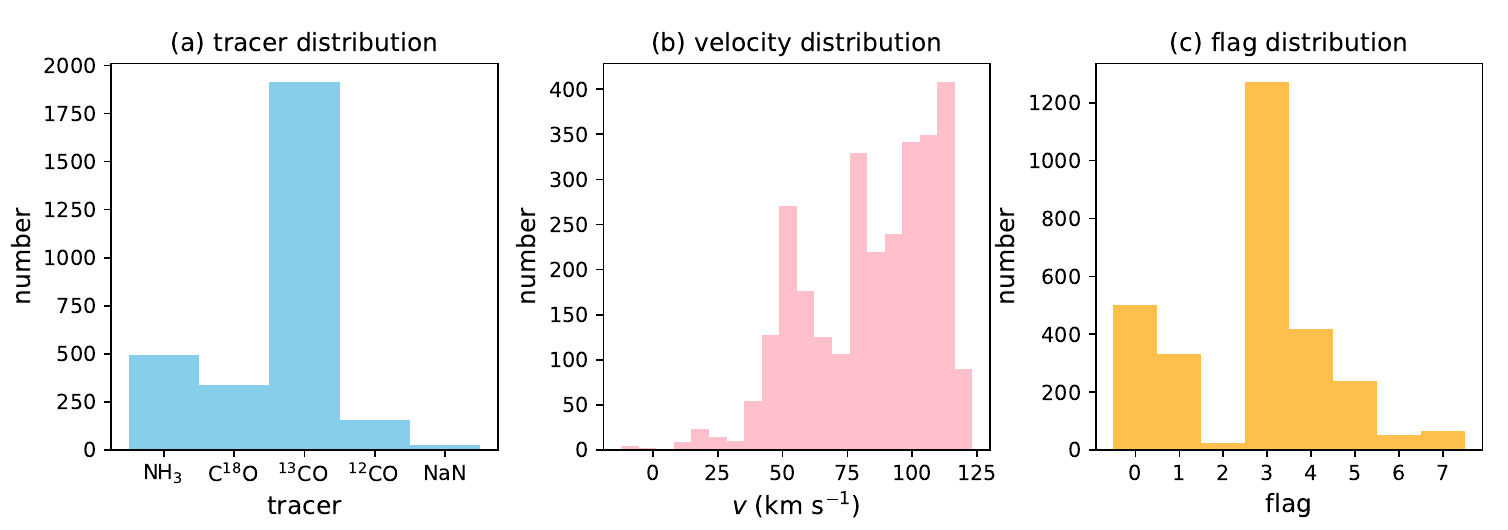}
\caption{ Velocity properties of the GASTON 1.15 mm leaves. {\bf (a):} Occurrences of the different tracers used for velocity assignments. The NaN column represents cases where no valid velocity was detected from any spectral line. {\bf (b):} Velocity distribution after trunk check. {\bf (c):}  Velocity flag distribution after trunk check. Flag 0 corresponds to NH$_3$, flags 1-2 correspond to C$^{18}$O, flags 3-5 correspond to $^{13}$CO, and flags 6-7 correspond to $^{12}$CO.}
\label{v_dis}
\end{figure*}

The reliability of the source velocity derived above decreases with increasing flag number. For example, the $^{12}$CO(1-0) velocities (with flags of 6 or 7) are the least reliable, while those from NH$_3$ (with a flag of 0) are the most reliable. We therefore decided to add another layer in our velocity identification scheme. We assumed that the dendrogram substructures of a given trunk (identified in $l, b$ space) should also appear as a coherent group in the position-position-velocity (i.e. $l, b, v$) space. So the velocities of all sub-structures belonging to the same trunk were checked. We consider sources with a flag $< 5$  having reliable velocities (see Appendix \ref{velocity_decision}) and therefore use their median velocity to adjust the velocities of other substructures with a flag of 5 or greater. However, this correction is only applied if the maximum velocity difference between these reliable sources is less than 10 \kms. 
%which means they are possibly clustered in velocity space and belong to the same cloud. 
If this conditions is met, the median velocity and maximum flag of those reliable sources are then passed on to substructures with a poor flag (i.e., 5, 6, 7) within this trunk. If not, the original velocity and flag remain unchanged. The updated velocity and flag are presented in column $v$ and flag in Tab. \ref{115table}. The distributions of the original velocity tracers, updated trunk velocity and updated trunk velocity flags for the 1.15 mm compact sources are plotted in Fig. \ref{v_dis}. Among those sources, $17\%$ have velocities derived from NH$_{3}$, $12\%$ from C$^{18}$O, $65\%$ from $^{13}$CO, and $5\%$ from $^{12}$CO. \textcolor{red}{}. Only $1\%$ of the compact sources do not have velocity measurements. Overall, $87\%$ of the GASTON-GP 1.15 mm have reliable velocities with flags smaller than 5. 

\subsection{Distance and radius}
Knowing the Galactic longitude, Galactic latitude, and radial velocity of each source, we can estimate the kinematic distance to the source using a Galactic rotation model. Here we use the Python wrapper for the Bayesian Distance Calculator ({\sc BD\_wrapper}) to solve the distance \citep{Riener2020}. The Bayesian Distance Calculator ({\sc BDC}) works based on the Galaxy model revealed by parallaxes of high-mass star formation regions from the Bar and Spiral Structure Legacy Survey (BeSSeL survey) \citep{Reid2016, Reid2019}. Sources inside the solar circle (that includes all GASTON-GP sources) have a kinematic distance ambiguity. Sources at two different distances along line-of-sight may have same line-of-sight velocity, referred to as the near and far   kinematic distances. {\sc BDC} can solve the ambiguity based on the assumptions made and spiral arm models used.

In {\sc BD\_wrapper} 2.4, we used the default parameters: {\verb|kd|} = 1, {\verb|sa|} = 0.85, {\verb|ps|} = 0.25, {\verb|gl|} = 1, {\verb|pm|} = None, which takes into account the kinematic distance, spiral arms, proximity to high-mass star-forming regions with parallax measurements, and the Galactic latitude value or displacement from the Galactic midplane. The numbers define the relative probabilities of the different priors. Once those parameters are set, BDC will give the near and far kinematic distance, and also the distance at the maximum of the likelihood distribution. Due to the high concentration of sources in the spiral arm when {\verb|sa|} is on, we decided not to blindly use the latter to avoid a potential spiral arm bias. Instead, we chose between the near and far kinematic distances based on which one was the closest to the solution provided by BDC.
%compared the maximum likelihood distance with the highest likelihood in the results with the distance of the tangent point ($\sim 7.5 $ kpc towards GASTON field). Based on the comparison, we chose the near or far kinematic distance to be assigned to the source.

Considering that maser is a good tracer to the star-forming region and some of them have accurate parallax measurements, we matched our catalogues with maser catalogue from the {\sc BDC} and adopted the parallax distances of masers \citep{Reid2019}. We compared the $(l,b)$ coordinates of the masers with the peak location of the GASTON-GP compact sources. If a maser falls within the radius of a compact source, we consider the maser source to be associated with it and adopt the maser distance. In total, 10 compact sources at 1.15 mm were found associated with maser sources, and 11 at 2.00 mm. These samples of 1.15 mm and 2.00 mm sources with maser association are fully cross-matched, as two 2.00 mm compact sources are cross-matched to the same 1.15 mm source. We replaced their distances using the maser distances accordingly. 

\begin{figure*}
\includegraphics[scale=0.58]{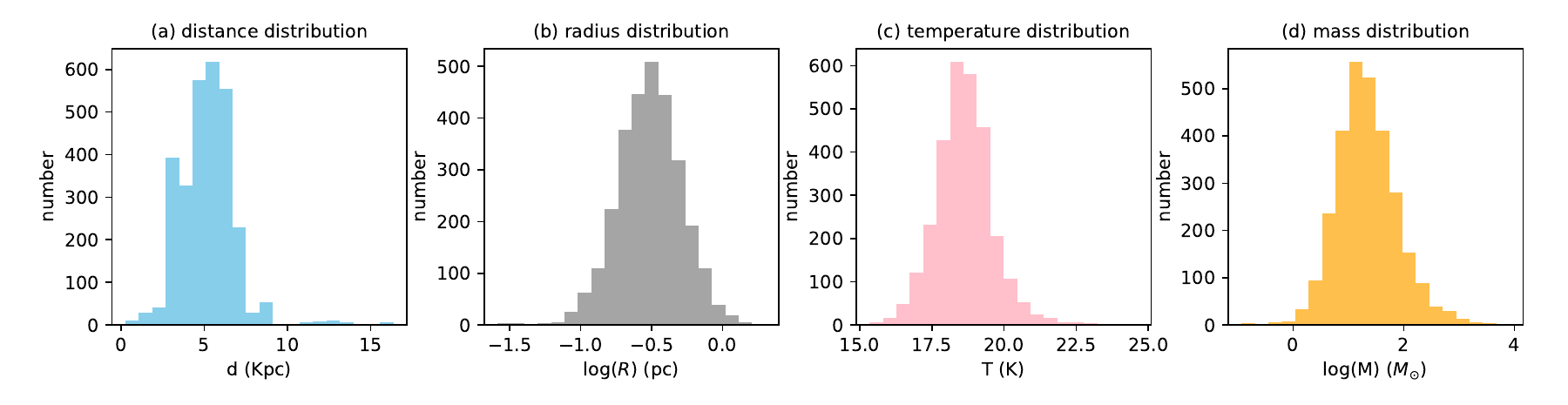}
\caption{Properties of the GASTON-GP 1.15 mm compact sources. {\bf (a):} Distance distribution.
{\bf (b):} Radii distribution. 
{\bf (c):} Temperature distribution.
{\bf (d):} Mass distribution.}
\label{para_dis}
\end{figure*}

Based on the derived distances and structure radii from {\sc Astrodendro}, we calculated the size of the compact sources: 
\begin{equation}
R = \sqrt{\frac{\Omega}{\pi}} d ,
\end{equation}
\noindent where $\Omega$ is the solid angle covered by the footprint of the source and $d$ is the kinematic distance we adopted. The distance and radius distribution of compact sources at 1.15 mm are shown in Fig. \ref{para_dis}. They have a median heliocentric distance of 5.20 kpc. The 16$^{\rm{th}}$ and 84$^{\rm{th}}$ percentiles is at 3.47 kpc and 6.57 kpc, respectively. For radii, they have a median value of 0.31 pc and a dispersion of 0.19 pc, with the 16$^{\rm{th}}$ and 84$^{\rm{th}}$ percentiles at 0.19 pc and 0.50 pc. The mean uncertainty on the kinematic distance is $29\%$, and therefore the same for the radii.

\subsection{Dust temperature}

To derive clump masses we need an estimation of the sources' dust temperatures ($T_{\rm dust}$). Here we use the colour temperature map from \cite{Peretto2016}, in which the temperature is calculated using the intensity ratio between 160 $\mu$m and 250 $\mu$m: 
\begin{equation} 
\frac{I_{160}}{I_{250}} = \frac{B_{160}(T_{\rm col})}{B_{250}(T_{\rm col})}\left(\frac{250}{160}\right)^{\beta},
\end{equation} 
where $I_{\nu}$ is the intensity in each pixel of the {\it Herschel} maps, and $\beta$ is the dust emissivity index. We adopted a mean Galactic plane value of $\beta = 1.8$ here \citep{Planck2011}.

 After matching the temperature map to the GASTON-GP source footprints, we derived the weighted mean temperature $T_{\rm{dust}}$ within the source footprints, where the weights are the GASTON-GP intensities on the filtered deep map. The dust temperature distribution of the GASTON-GP 1.15 mm compact sources is plotted in Fig. \ref{para_dis}. They have a temperature range from 15.3 to 24.6 K, with a mean value of 18.6 K. The {\it Herschel} calibration uncertainty translate into a colour temperature uncertainty of $\sim1.0$~K. However, this uncertainty does not take into account systematic uncertainties resulting from, e.g., a single line-of-sight temperature. For sources with lower (larger) mass-averaged temperatures than that of the background, the temperatures we use here will be overestimated (underestimated) \citep{Peretto2010}.

\subsection{Mass}
The masses of compact sources are calculated from their integrated flux density $S_{\nu}$, dust temperature $T_{\rm{dust}}$ and distance $d$ via the following equation (assuming optically thin dust emission):
\begin{equation} 
M = \frac{S_{\nu}\ d^2}{\kappa_{\nu}\ B_{\nu}(T_{\rm col})},
\end{equation} 

\noindent where $\kappa_{\nu}$ is the dust emissivity. $\kappa_{\nu}$ is calculated by $\kappa_{\nu} = 0.1(\nu/1{\rm THz})^{\beta}$ \citep{Beckwith1990}. With $\beta = 1.8$, $\kappa_{260 {\rm GHz}} = 0.009\ {\rm cm}^2\ {\rm g}^{-1}$ at 1.15 mm and $\kappa_{150 {\rm GHz}} = 0.003\ {\rm cm}^2\ {\rm g}^{-1}$ at 2.00 mm.  The dust emissivity values include a dust-to-gas mass ratio of 1\%.

The distribution of clump masses is plotted in Fig. \ref{para_dis}. There is a wide range of masses from 0.014 to $8.7\times 10^3$ \sm, with a median mass of 20 \sm, 16$^{\rm{th}}$ and 84$^{\rm{th}}$ percentiles at 7 and 72 \sm. At the low-mas end, 14 sources have masses smaller than 1 \sm, which turns out to be faint and nearby ($d < 1.5$ kpc) sources. At the high-mass end, 324 sources have masses larger than 100 \sm. Overall, the mass distribution is clearly dominated by rather low-mass sources. This is a direct reflection of the high sensitivity of the GASTON-GP observations. 

All physical properties of the GASTON-GP compact sources were derived independently at 1.15 mm and 2.00 mm. Fig. \ref{v_dis} and Fig. \ref{para_dis}, and Tab. \ref{115table} present the results obtained at 1.15 mm, while the corresponding properties, and related uncertainties\footnote{Mass uncertainties were derived through Monte Carlo propagation by combining the normally distributed uncertainties on the flux density, kinematic distance and temperature. A factor of 2 in the systematic uncertainty of the dust emissivity, $\kappa_{\nu}$, was not included in our uncertainty estimation.}, derived at both wavelengths are available in the online catalogues .

\section{Compact source variability} \label{variability study}

The key scientific angle of this study is the millimetre variability of massive protostellar sources. However, the GASTON-GP survey was not designed for variability study. Varying data quality across runs can be seen from Fig. \ref{115_11_examples}. As a result, a dedicated treatment of these data-quality-related variations across the different maps had to be developed, which we present in this section.

\subsection{Relative calibration}\label{flux calibration}
As mentioned in Sec.~\ref{data reduction}, the absolute flux calibration uncertainty of the GASTON-GP data is $\sim14\%$ at 1.15 mm and $\sim7\%$ at 2.00 mm. A significant fraction of this uncertainty is driven by differences in the run-to-run observing conditions, which generate large systematic flux variations from one run to the next. In the following, we describe the method we designed and implemented to achieve a higher level of relative calibration uncertainty. 

First, we made the assumption that only a small number of GASTON-GP sources might experience variability during the 4-year period of the observations, and therefore GASTON-GP sources should have a constant flux throughout that period. Then, we select a sample of bright high-SNR sources throughout the GASTON-GP field and across all runs to serve as {\it calibration sources}. We use the run-to-run statistical variations of this sample to perform a relative calibration of all GASTON-GP sources allowing thus to correct for varying data quality.  In the following, we describe the method in more details. 

\begin{figure*}
\includegraphics[scale=0.85]{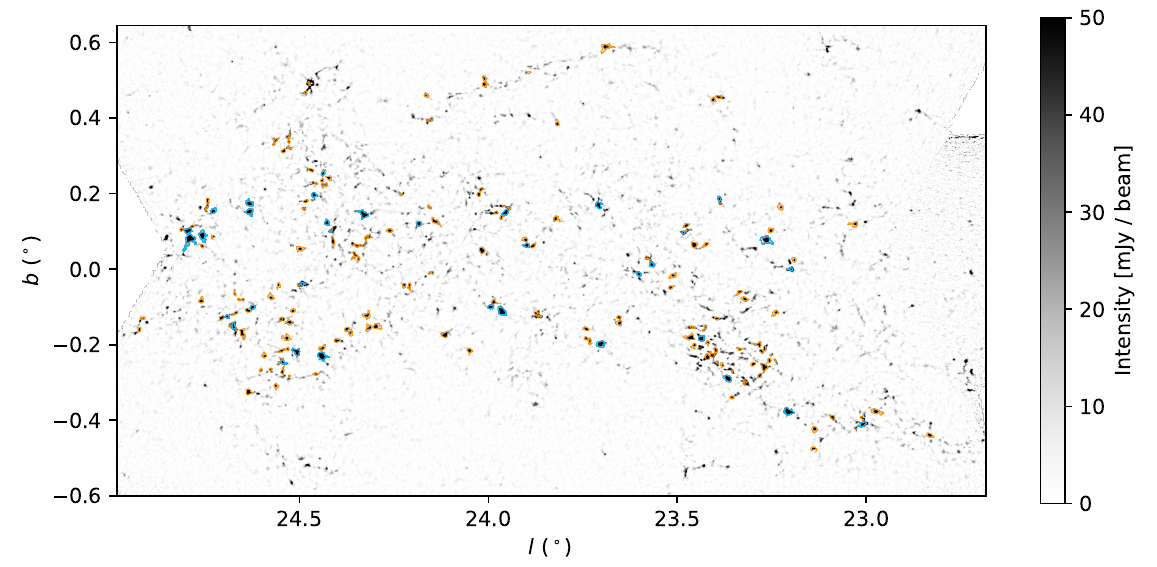}
\caption{Footprints of the 36 calibration sources (blue contour) and 206 high-SNR sources (orange contour) at 1.15 mm. The background image is the 1.15 mm filtered deep map.}
\label{compact_sources}
\end{figure*}

\vspace{0.1cm}
\noindent \textbf{Step 1: Calibration source selection.} \\
For each compact source $i$, we collected its peak intensity $(I_{i,j})$ at run $j$ and corresponding signal-to-noise ratio ${\rm SNR}_{i,j}$ at the peak location from the filtered images. Considering that the intensity of the fainter source is more affected by noise, we selected a group of calibration sources with the smallest S/N ratio above 20 and with effective flux measurements available in more than 8 runs. In total, 36 compact sources at 1.15 mm and 37 compact sources at 2.00 mm were selected as calibration sources. The footprints of these calibration sources are shown in Fig. \ref{compact_sources}.

\begin{figure*}
\includegraphics[scale=0.8]{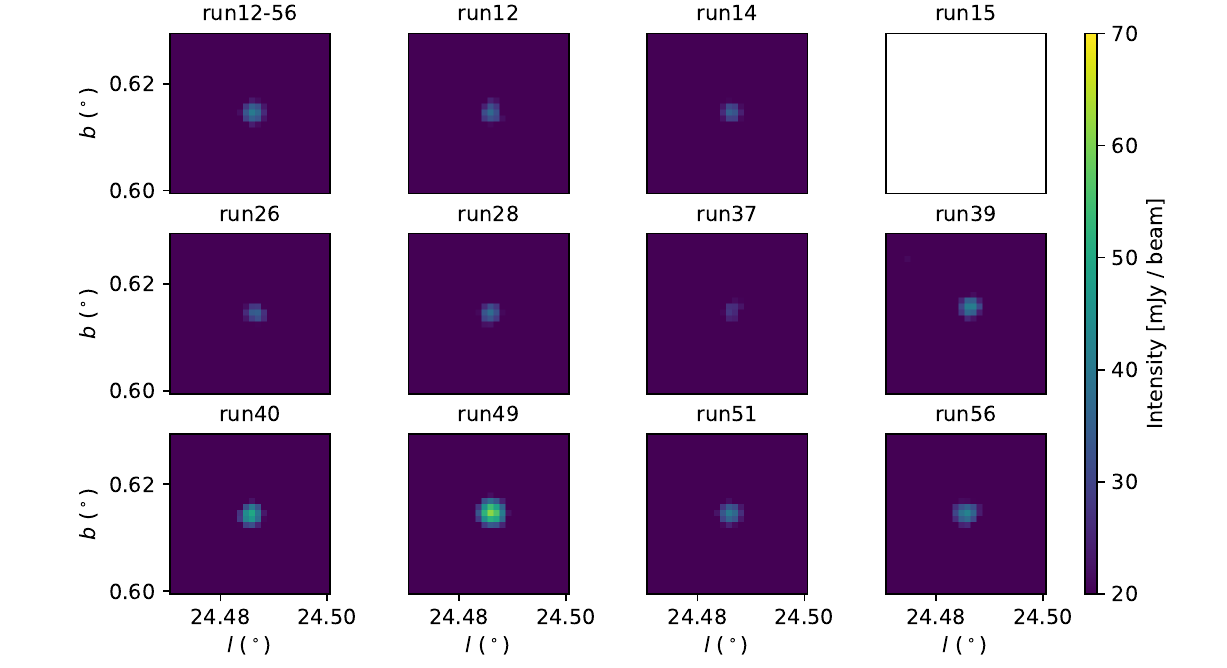}
\caption{The 12 subplots show the the calibrated emission for the 2.00 mm source S174, i.e. the only robust variable candidate of the GASTON-GP sample. Data maps from Run 12 to Run 56 have been calibrated by applying the flattening parameter (See Sec. \ref{flux calibration}). There is no observation available in Run 15 for this source.}
\label{compact_example}
\end{figure*}

\vspace{0.1cm}
\noindent \textbf{Step 2: High-SNR source selection.}\\
Compact sources are extracted from the deep SNR map (see Sec. \ref{method}). However, due to much lower integration time and different coverage of individual runs, not all the sources are visible or of high-quality across all runs. We thus compile a sample of high-SNR sources that we can use for variability study. We chose compact sources with a SNR $>5$ in at least 8 of the 11 runs. % Note that this criterion and the one for calibration sources include all the 11 data runs (see Fig. \ref{115_11_examples}). 
The number of high-SNR sources at 1.15 mm and 2.00 mm is 206 and 175, respectively. The footprints of the high-SNR 1.15 mm sources are shown in Fig. \ref{compact_sources}. Cutout images  have been made so that visual inspection of each source can be performed (see Fig.~\ref{compact_example} for one example of 2.00 mm high-SNR source). Our variability study focuses on these high-SNR sources only.

\vspace{0.1cm}
\noindent \textbf{Step 3: Relative calibration estimation.} \\
In addition to the run-to-run calibration variations mentioned above, scan-to-scan data quality variations can clearly be seen across the different runs (see Fig.~\ref{115_11_examples}). As a result, our relative calibration scheme must attempt to capture such local variations. To achieve this, for each calibration source, we calculate the ratio $f_{0,i,j}$ of their intensity $I_{i,j}$ measured at a given run to the intensity $I_{i}$ measured in the deep map:
\begin{equation}
f_{0,i,j} = \frac{I_{i,j}}{I_i}.
\end{equation}

\noindent We call $f$ the flattening parameter. We then assume that the data quality is locally similar on a scale of a scan width, i.e. 0.6 degree. Since the scanning direction is mostly along the Galactic latitude axis, we defined around each high-SNR source a $0.6^{\rm{o}}$-wide rectangular slice of the GASTON-GP field, then selected all calibration sources falling within it, and calculated the resulting median flattening parameter $f_{c,i,j}$. Following then the assumption that most calibration sources have flat light curves, the calibrated peak intensity $I_{c, i, j}$ for high-SNR source $i$ in run $j$ is given by: 
\begin{equation}
I_{c,i,j} = \frac{I _{i,j}} {f_{c,i,j}}.
\end{equation}

\noindent The calibrated peak intensity $I_{c,i,j}$ of each high-SNR source was then used for the variability study. The same procedure was applied at both wavelengths separately.

\subsection{Uncertainty calculation on calibrated peak intensities}
The uncertainty on the calibrated peak intensities for each high-SNR source comes from two parts: one is the intrinsic Poisson uncertainty on the observed intensities $I_{i,j}$, and the other is the uncertainty introduced by the median flattening parameter $f_{c,i,j}$. We used Monte Carlo sampling to derive the combined uncertainty.

Regarding the Poisson uncertainty on the intensity, we used the uncertainty from the RMS map. We assumed that the uncertainty of the intensity follows a normal distribution, and for each source, we used the peak intensity as the mean of the distribution and the rms noise at the peak location as its standard deviation. Regarding the uncertainty on the median flattening parameter $f_{c,i,j}$, we used the distribution of flattening parameters $f_{0,i,j}$ of the calibration sources used for each high-SNR source. Then, for each source $i$ at run $j$, we produced a distribution of calibrated peak intensities by randomly selecting a peak intensity from the normal peak intensity distribution and a flattening parameter from the flattening parameter distribution. We iterated this procedure 1000 times and then took the difference between the 16$^{\rm{th}}$ and 50$^{\rm{th}}$ percentiles of the resulting calibrated intensity distribution as the lower uncertainty $I_{c,i,j}^{{\rm lower}}\_{\rm err}$ and the difference between the 84$^{\rm{th}}$ and 50$^{\rm{th}}$ percentiles as the upper uncertainty $I_{c,i, j}^{{\rm upper}}\_{\rm err}$.

\subsection{Validation of the method}
To assess the quality of our relative calibration procedure on a run-by-run basis, and check whether systematic residuals can impact our variability study, we compared the calibrated intensities of high-SNR sources with their reference intensity from the deep map.

\begin{figure*}
\includegraphics[scale=0.8]{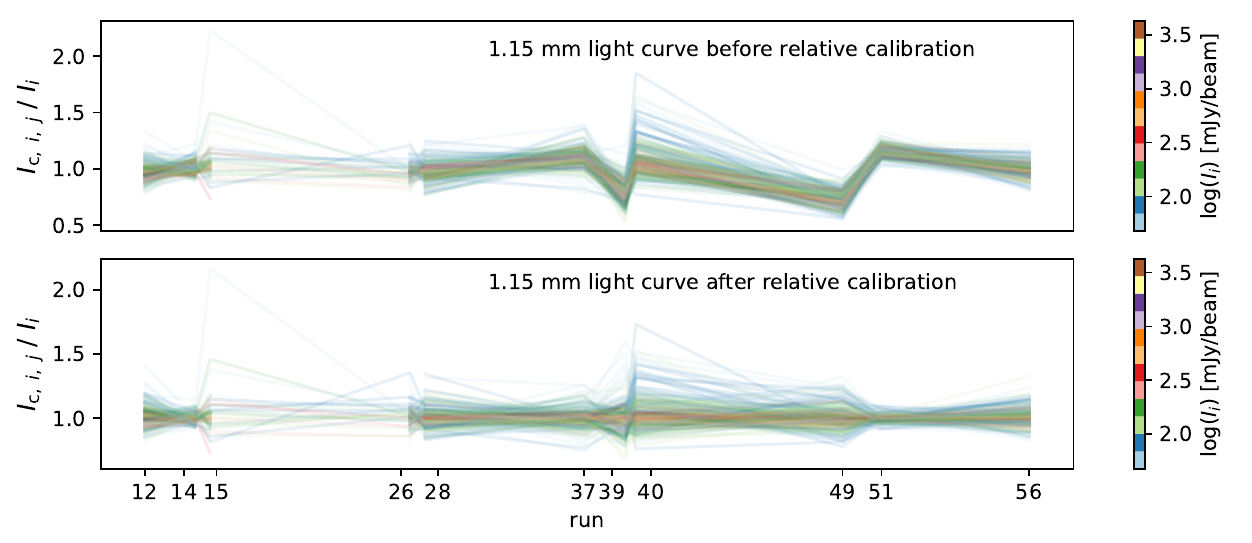}
\caption{Normalised light curves of 1.15 mm high-SNR sources before ({\bf top}) and after ({\bf bottom}) relative calibration. Light curves are colour-coded with respect to their reference peak intensities. }
\label{115lc}
\end{figure*}

We first checked the light curves of high-SNR sources before and after relative calibration. Those at 1.15 mm are plotted in Fig. \ref{115lc}. Before relative calibration, many sources exhibit similar trends at certain runs, with peaks at run 40 or run 51 or dips appearing at run 39 or run 49, indicating the presence of systematic variations in the data. After applying the method, these common trends are largely removed, and most light curves become flatter and more randomly distributed around the reference intensities. Figure \ref{115lc} also shows that sources with larger apparent post-calibration variations are generally the faintest sources whose intensity measurements are dominated by Poisson noise.  

\begin{figure*}
\includegraphics[scale=0.65]{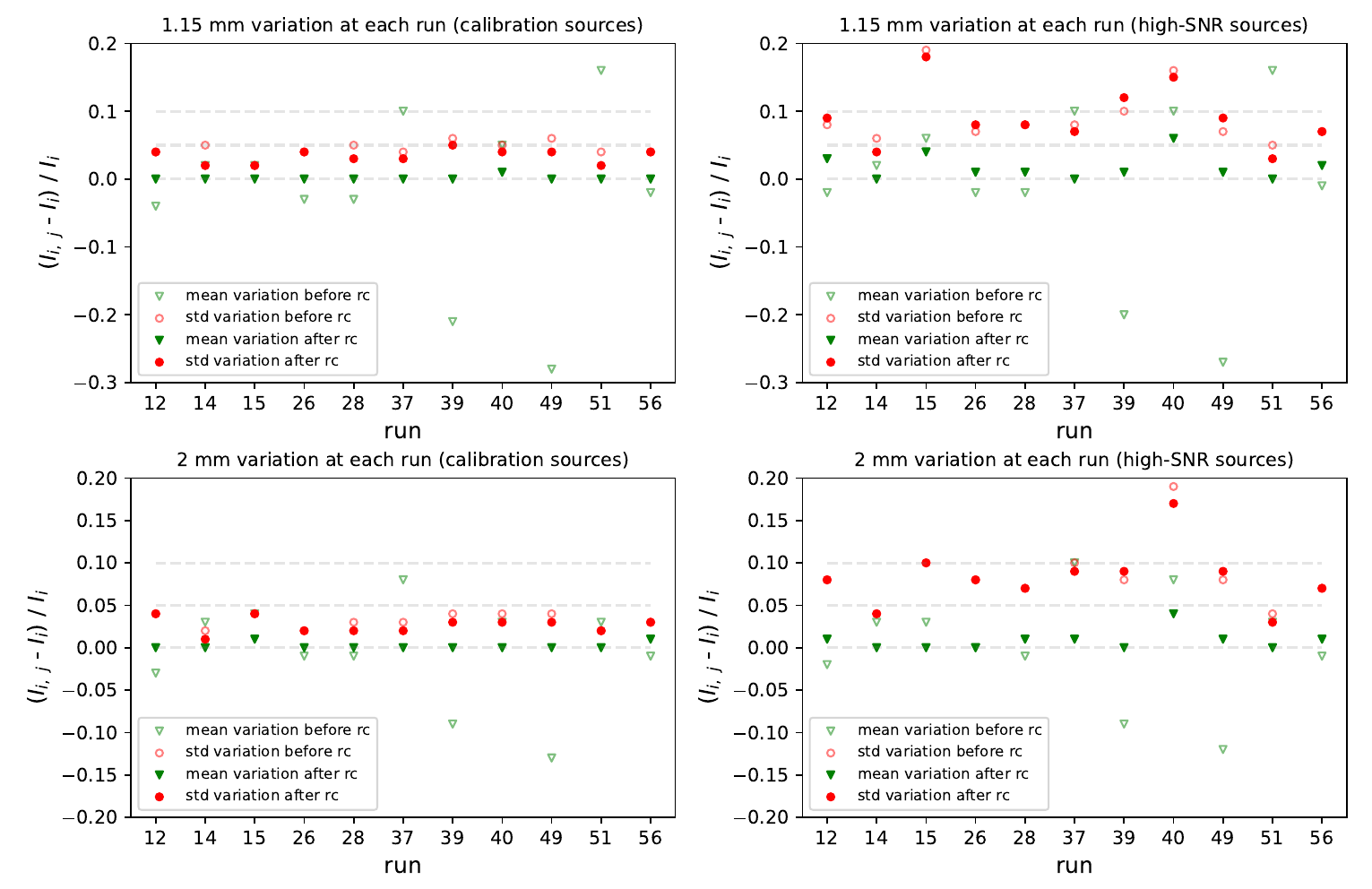}
\caption{Mean (triangle) and standard deviation (circle) of the relative intensity variations $\alpha$ before (empty) and after (filled) relative calibration at 1.15 mm ({\bf top}) and 2.00 mm ({\bf bottom}). {\bf Left} panels are for calibration sources and {\bf right} panels are for all the high-SNR sources. In the legend, std refers to the standard deviation and rc refers to the relative calibration. The run numbers indicated in each panel corresponds to $j$ in Eq.~(\ref{alphar2r}). The horizontal lines correspond to $\alpha=0$, $\alpha=0.05$, and $\alpha=0.1$. }
\label{run_to_run_variation}
\end{figure*}

Following this qualitative validation, we quantitatively assess the calibration performance.
For each source, we calculated the ratio $\alpha$ following:
\begin{equation}
\alpha = \frac{I_{c,i,j} - I_{i}}{I_{i}},
\label{alphar2r}
\end{equation}
which quantifies the relative post-calibration intensity variation at each run. If most of the sources have flat light curves, the mean of $\alpha$ values for all sources identified in run $j$ should be around 0, while the standard deviation of calibration sources should give use the achieved relative intensity calibration uncertainty. The mean and standard deviation of $\alpha$ for each run are shown in Fig. \ref{run_to_run_variation}, for both calibration sources and high-SNR sources, and for both pre- and post-calibration cases. Focussing first on calibration sources, one can see that the mean $\alpha$ value has significantly changed for most runs, and now align, within a couple of \% with the $\alpha=0$ line. This is a sign that we have effectively corrected the run-to-run absolute calibration variations. We also notice that the standard deviation of $\alpha$ has not changed much, as expected, and is overall $<5\%$ at both wavelength. In fact, the average standard deviations among calibration sources across 11 runs of 3.4\% at 1.15 mm and 2.6\% at 2.00 mm can be taken as a measure of the relative intensity calibration uncertainty, a factor of 3 to 4 better than the absolute intensity calibration uncertainty. 

If we now focus on the high-SNR sources, Fig. \ref{run_to_run_variation} shows that, even though similar trends are achieved, the variations are larger, especially for specific runs, highlighting the varying quality of each individual runs and their corresponding noise levels (see Tab. \ref{run-infor}), but also the different number of high-SNR sources available in each run. 

Overall, our relative calibration scheme performs well, effectively removing most of the run-to-run absolute calibration variations. Residual systematics remain, in particular in some specific problematic runs (runs 15, 26, 40) but those are taken into account in the variability analysis that follows. 

%The number of calibration sources used to estimate the median flattening parameter varies depending on the the location of the high-SNR sources. This affects the robustness of the relative calibration. For 95\% of the high-SNR sources at least 6 calibration sources have been used across all runs, except fewer source in the corner of the map. From the $\alpha$ distribution mentioned above, we believe the relative calibration method characterized the global variation at each runs well and effectively removes run-to-run systematic offsets. Note the variation dominated by the random noises for faint sources can not be modelled and the complicated variation in the environment can not be fully calibrated. 

\subsection{Light curves}\label{light curve sec}

To visualize the intensity change for each high-SNR source, we plotted their light curves over time at both 1.15 and 2.00 mm. Some light curve examples can be found in Fig. \ref{115lc} and Fig. \ref{candi_lc}. 

\begin{figure*}
\includegraphics[scale=0.7]{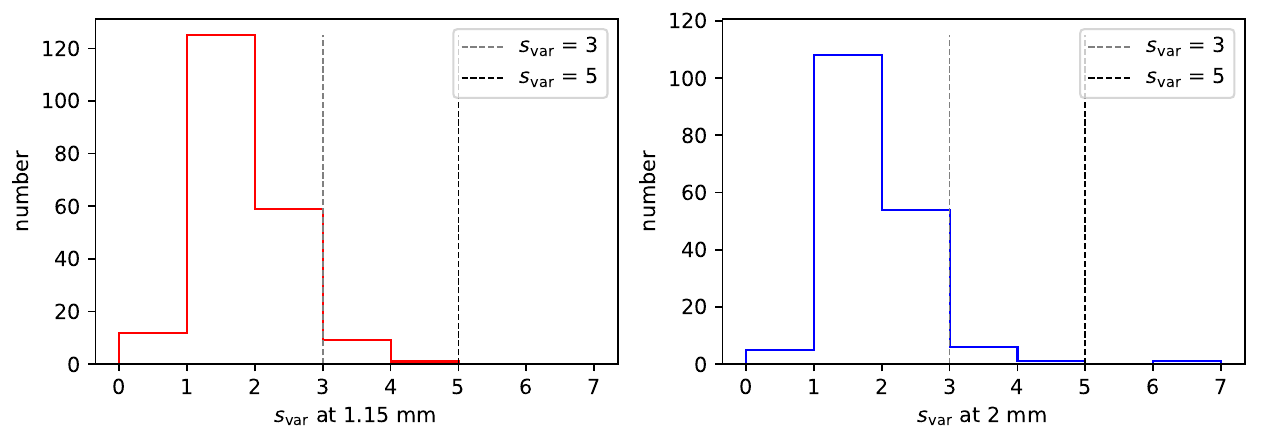}
\caption{$s_{{\rm var}}$ distribution at 1.15 mm ({\bf left}) and at 2.00 m ({\bf right}). The grey and black dashed lines indicate $s_{\rm var}=3$ and $s_{\rm var}=5$, respectively.}
\label{svar_dis}
\end{figure*}

To characterise the intensity variation of GASTON-GP high-SNR sources, we defined and calculated the significance of a source's variability, $s_{{\rm var},\ i}$, by taking its maximum intensity deviation from the reference intensity and divided the difference over the calibrated intensity uncertainty: 
\begin{equation}
s_{{\rm var},i}={\rm max}_j\left(\frac{|I_{c,i,j}-I_i|}{I_{c,i,j}\_{\rm err}}\right), 
\end{equation}
where we used $I_{c,i,j}^{{\rm lower}}\_{\rm err}$ when $I_{c,i,j} > I_{i}$ and $I_{c,i,j}^{{\rm upper}}\_{\rm err}$ when $I_{c,i,j} < I_{i}$.
%A high-SNR source shows variation signals when $s_{{\rm var},\ i} > 3$. 
The distributions of $s_{{\rm var}}$ at both wavelengths are shown in Fig. \ref{svar_dis}.
Among 206 high-SNR sources at 1.15 mm, we identified 10 sources with $s_{{\rm var},i} > 3$, and none with $s_{{\rm var},i} > 5$. At 2.00 mm, we found 8 of the 175 high-SNR sources with $s_{{\rm var},i} > 3$, and 1 with $s_{{\rm var},i} > 5$.

Accretion bursts drive luminosity changes across all wavelengths \citep[e.g.][]{Johnstone2013}. Therefore, the variable trend observed at one of the GASTON-GP wavelengths should also be observed at the other. This correlation at different wavelengths could be used to confirm the existence of variable sources with low $s_{{\rm var},i}$ value.  
To check the light curves at both wavelengths, we first matched the high-SNR sources at 1.15 mm and 2.00 mm by checking whether the peak pixel at one wavelength fell onto the footprint at the other wavelength. Among the 206 high-SNR sources at 1.15 mm, 139 of them have high-SNR counterparts at 2.00 mm.
For the matched pairs, we analysed the Pearson correlation coefficient and p-value between the 1.15 mm and 2.00 mm light curves. The null hypothesis assumes no correlation between the two light curves. If the p-value is below 0.05, we reject the null hypothesis and consider the two light curves correlated, indicating a similar variable trend at both wavelengths. Overall, 15 light curve pairs exhibit p-values below 0.05 among the 139 matched sources, suggesting a strong correlation between their 1.15 mm and 2.00 mm light curves.

%% file: tex/4_results.tex
\subsection{Variability results}
\subsubsection{Variable and correlated light curves}\label{candidates}
\begin{figure*}
\includegraphics[scale=0.58]{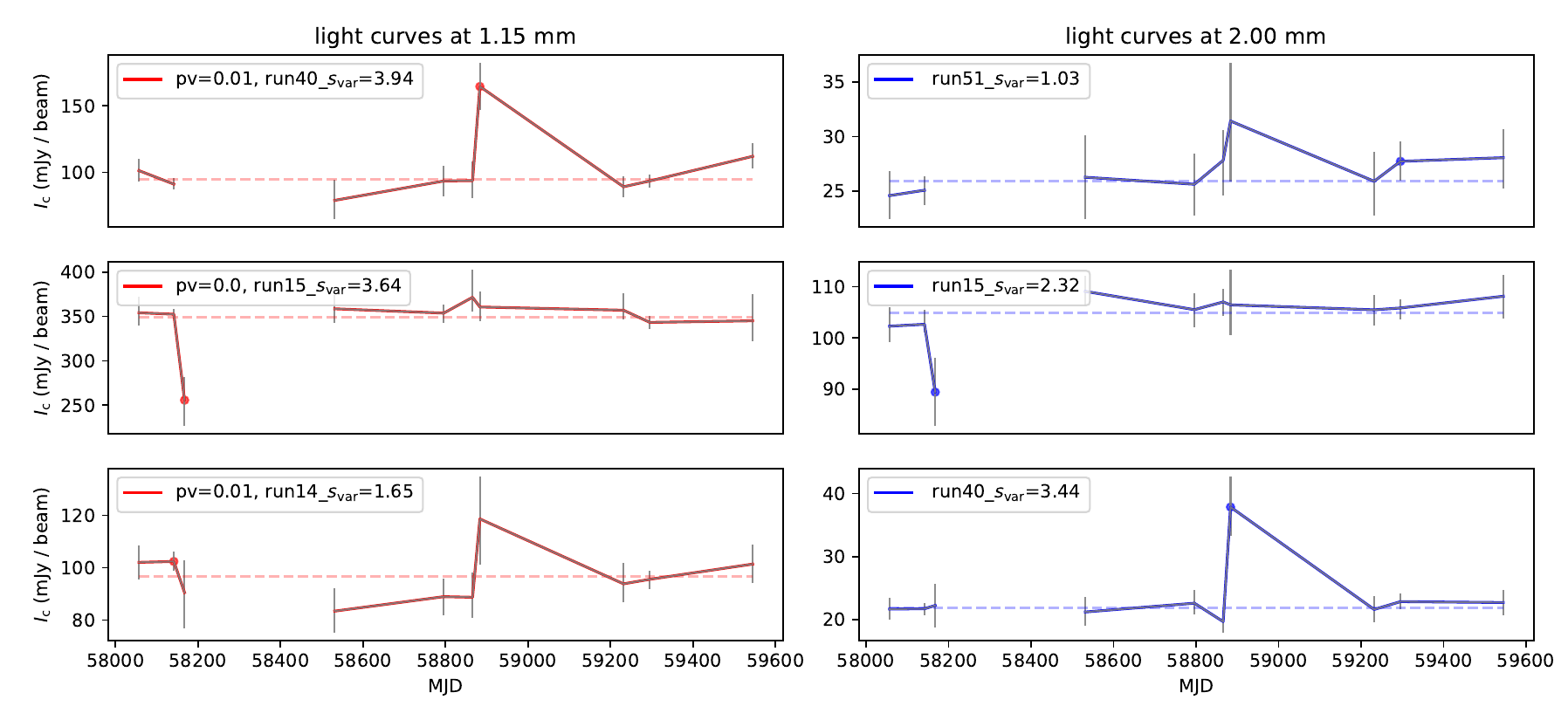}
\caption{Light curves of 3 variable candidates: G23.139-0.474 ({\bf top}), G23.420-0.234 ({\bf middle}), G23.573+0.025 ({\bf bottom}). The left panels shows light curves at 1.15 mm, and the right panels those at 2.00 mm for the same sources. The dashed line in each subplot marks the reference flux of each source from the filtered deep maps. We present the significance of the maximum variation $s_{\rm var}$ and the p value from the Pearson correlation study in the legend. The point in each plot marks the run where $s_{\rm var}$ occurs.}
\label{candi_lc}
\end{figure*}

Based on the two parameters $s_{\rm var}$ and p-values, we consider a source as a variable candidate if it satisfies the following two conditions: $s_{\rm{var}}\ge3$ at least at one wavelength and p-value is below 0.05. 
%, which means it is variable at least at one wavelength and has similar variable trend at both bands. 
Under those two conditions, we end up with 3 variable candidates: G23.139-0.474, G23.420-0.234, G23.573+0.025. 
%They all have $s_{\rm{var}}\ge3$ at only one wavelengths. 
Their light curves are plotted in Fig. \ref{candi_lc}. One can clearly notice the similarities between those different light curves, with peaks in run 40 and drops in run 15, two runs that are known to be of poorer quality. These similarities are clearly driven by the residuals of our run-to-run relative calibrations. These three sources are therefore unreliable variable candidates.

%Among them, G23.139-0.474 and G23.573+0.025 show intensity increase at run 40, which is known to be the run with poor data quality, making them very spurious. Their $s_{\rm var}$ values are also below 4, indicating that the variability signals are not strong. The other source, G23.420-0.234, was found to lie near the edge of the map at the variable run 15, where the intensity is less reliable. Given these issues, we do not consider any of these sources to be variable candidates.

\subsubsection{A robust 2.00 mm variable compact source G024.485+0.614}
\begin{figure}
\includegraphics[scale=0.58]{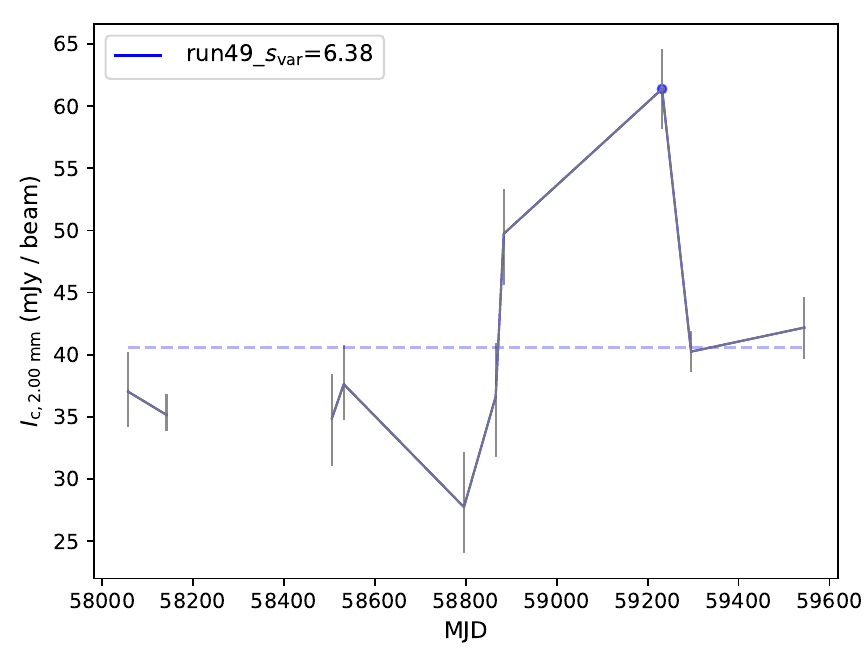}
\caption{2.00 mm light curve for the 2.00 mm variable candidate G024.485+0.614 with $s_{{\rm var}} = 6.38$. The point marks the run where $s_{\rm var}$ occurs.}
\label{2mmcan}
\end{figure}

Although no other sources satisfy the set of criteria listed above, one compact 2.00 mm source is highly variable with $s_{\rm var} = 6.38$. Its light curve (Fig. \ref{2mmcan}) displays a strong brightening event peaking at run 49 (see also Fig. \ref{compact_example}). Notably, its pre-calibration light curve also shows a variability peak while all calibration sources show the opposite trend, further supporting the variability. Located at ($l=24.485^{\rm{o}}$, $b=0.614^{\rm{o}}$), this source is isolated from any star-forming clump. It has peak intensities of 40.6 ${\rm mJy\,beam^{-1}}$ at 2.00 mm, and 19.9 ${\rm mJy\,beam^{-1}}$ at 1.15 mm, and has also been detected as a point-source with flux densities between $\sim50$ to $100$~mJy in the radio (e.g. 1.4 GHz VLA survey - \cite{Zoonematkermani1990}, 3 GHz VLASS survey - \cite{Gordon2021}, 5 GHz CORNISH survey - \cite{Purcell2013}, 1.3 GHz SMGPS survey - \cite{Mutale2025}). 
% SARAO MeerKAT 1.3 GHz Galactic Plane Survey (SMGPS) with a peak intensity of 99.98 mJy/beam \citep{Mutale2025}. 
The negative spectral index of its spectral energy distribution, combined with the lack of any surrounding dusty clumps, clearly shows that this variable source is not a protostellar object. Its brightness also excludes the possibility of a high-redshift galaxy \citep{Bethermin2025}. The remaining option could be that this variable sources is a Galactic planetary nebulae. However, the lack of mid-infrared emission in {\it Spitzer} data sheds doubt on the planetary nebulae hypothesis.  Although its physical nature is unclear, the 50\% variability of this source makes it an interesting target for follow-up investigation.

\subsubsection{GASTON-GP light curves for published variables} \label{variable_comparison}

The GASTON-GP survey mapped an area of the Galactic plane that has been extensively studied over the past 30 years or so. As a result, a large data bank is available for this region. This allows us to examine how the millimetre light curves from GASTON-GP survey behave for known variable sources that have exhibited burst-like events at other wavelengths.

Based on previous NEOWISE mid-infrared observations and maser observations, the source G24.32+0.14 showed a 1.3 mag brightness increase in the W2 (4.6 $\mu$m) band during September 2019, with a methanol maser flare detected at similar time \citep{Hirota2022}. This event occurred within the time span covered by the GASTON-GP survey, enabling a direct comparison between infrared and millimetre variability.

From another work using 10-year NEOWISE observation and ATLASGAL dataset, \citet{Lu2024} has conducted a study on the near infrared variability of ATLASGAL sources. Based on the period and pattern of the light curves, they classified the variables into secular (linear, sin, sin+linear) and stochastic variables (burst, drop, and irregular). Sources with stochastic luminosity bursts are likely undergoing accretion bursts. After matching the coordinates, two of the NEOWISE stochastic burst sources were observed within $11''$ (the GASTON-GP beam at 1.15 mm) of a GASTON-GP source: G23.8175+0.3841 and G23.2433-0.2392. 

We collected the NEOWISE data of these sources from the NASA/IPAC Infrared Science Archive. For each source, we used its peak location and crossed match with NEOWISE-R Single Exposure (L1b) Source Table using a search radius of 11 arcsec. As the radius is almost twice larger than NEOWISE resolution, several NEOWISE sources might satisfy our matching criterion. In this situation, we kept the one closest to the GASTON peak location. During each epoch, multiple observation measurements are made for each NEOWISE source. We thus averaged all measurements taken within 10-day time window for each source in each epoch. Only W2 (4.6 $\mu$m) band is used here.   

\begin{figure*}
\includegraphics[scale=0.5]{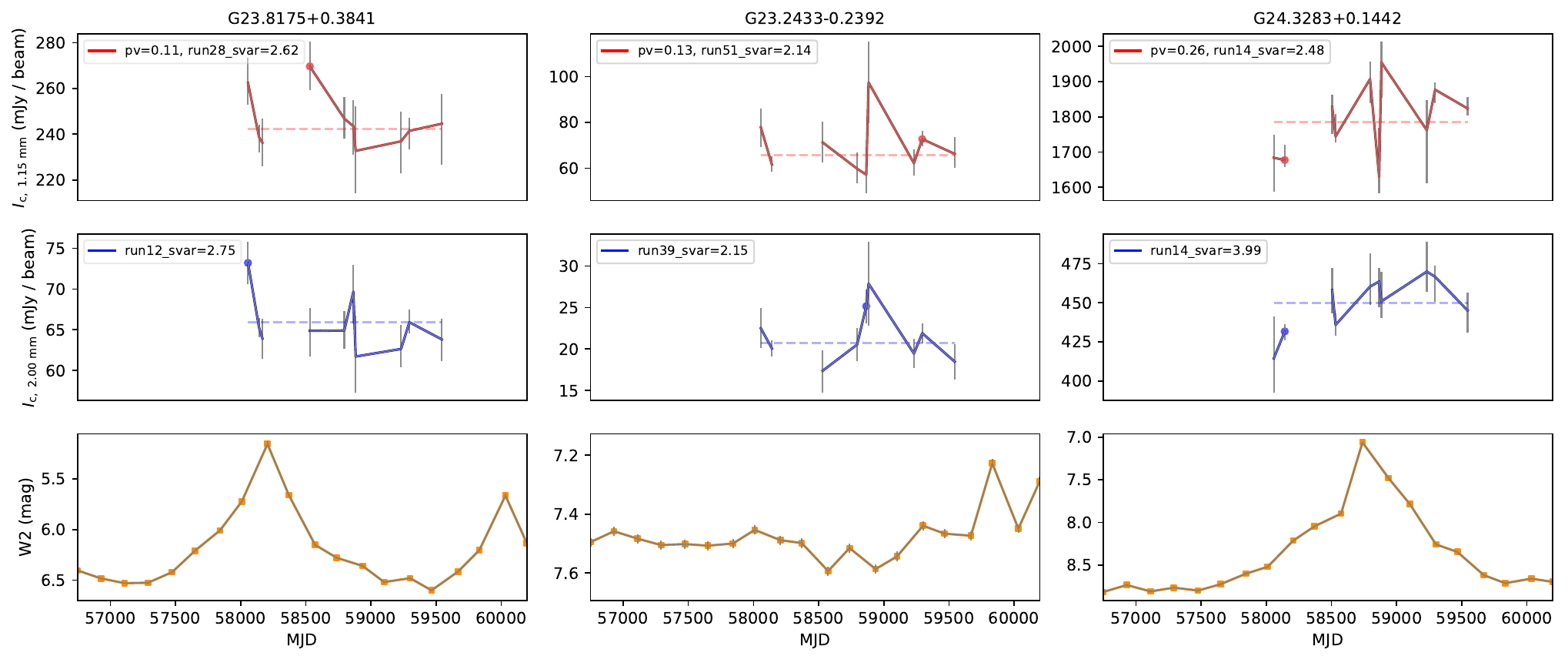}
\caption{Light curves for three sources with stochastic bursts in the NEOWISE infrared data: GASTON-GP 1.15 mm light curves ({\bf top, red}), GASTON-GP 2.00 mm light curves ({\bf middle, blue}), NEOWISE W2 (4.6 $\mu$m) light curves ({\bf bottom, orange}). The p-value between 1.15 mm and 2.00 mm light curves and $s_{\rm var}$ are mentioned in the legend at first two rows. The point in each GASTON light curve marks the run where $s_{\rm var}$ occurs.}
\label{NEOWISE}
\end{figure*}

To further investigate the millimetre variability of these three known variable sources, we compared their GASTON-GP light curves to the NEOWISE W2 band data in Fig. \ref{NEOWISE}. As the infrared emission comes from the inner disc and millimetre emission traces the outer envelope after reprocessing the radiation from the core, millimetre variability is expected to be delayed after the infrared burst, which can also be changed by the beam size \citep{Johnstone2013, Francis2022}. 
In Fig. \ref{NEOWISE}, GASTON-GP light curves exhibit different trends from the corresponding NEOWISE light curves. We can not match the variations between them. Although a significant flux increase is observed for G24.32+0.14 between MJD 58000 and 58300 at both 1.15 and 2.00 mm, which appears broadly consistent with the NEOWISE light curve at the similar time, the remaining of the GASTON-GP light curve is too noisy to confirm the existence of matching millimetre burst.
Consequently, within the uncertainties of our data, no significant correlation is observed between the millimetre and the infrared light curves. This highlights the observational challenges in detecting variability at millimetre wavelengths.

\subsubsection{Potential variable candidates with $s_{\rm var} > 3$}
In addition to the sources discussed above, we identified another 12 sources with moderate variability ($3 < s_{\rm var} < 5$) at either 1.15 mm or 2.00 mm.
Since these variations are not correlated at both wavelengths we cannot confidently determine whether the intensity changes are intrinsic.
These sources are listed in Tab. \ref{potential_sources}. Although they do not meet our criteria for robust variability, they may represent potential candidates for further study. Future observations with improved calibration will be needed to establish whether these sources exhibit genuine variability.

\begin{table}
\caption{
Information for the potential variable candidates with $3< s_{\mathrm{var}} < 5$.  
Columns: (1) source ID; (2) $s_{\mathrm{var}}$ value; (3) wavelength showing variability;  
(4) observing run in which the variation occurs; (5) variation trend.
}
\label{potential_sources}
\begin{tabular}{ccccc}
\hline
source id&  $s_{\rm var}$  & wavelength & run & variation  \\ \hline
G023.206-0.378 & 4.43 & 1.15 mm & run28 &  drop \\
G023.409-0.228   & 3.69  & 1.15 mm  &  run37 & drop \\
G023.656-0.128  & 3.28  & 1.15 mm  & run15  & peak \\
G023.031+0.120   &  3.65 & 1.15 mm  &  run56 &  peak\\
G024.633+0.152   &  3.29 & 1.15 mm  & run56  & peak \\
G023.943+0.159   &  3.40 & 1.15 mm  & run37  & peak \\
G023.958+0.167   &  3.28 & 1.15 mm  & run39  & drop \\
G024.473+0.487   & 3.64  & 1.15 mm  &  run26 & drop \\
G023.091-0.391   &  3.14 & 2.00 mm  & run56  & drop \\
G023.300-0.251   &   3.79 & 2.00 mm  &  run56 & drop \\
G024.329+0.144   & 3.99  & 2.00 mm  &  run14 & drop \\
G024.530+0.323   &  4.44 & 2.00 mm  & run37  &  peak \\  \hline
\end{tabular}
\end{table}

%% file: tex/5_discussion.tex
\section{Discussion}  \label{discussion}

Our variability analysis revealed no robust detections of bursting protostellar sources within the GASTON-GP field. The absence of significant variability raises the question of whether such events are intrinsically rare or simply difficult to detect at millimetre wavelengths. In the following section, we estimate the expected number of bursts in GASTON-GP region and discuss about the limitations of the GASTON-GP observations for variability studies.

\subsection{Expected number of variable sources in the GASTON field} \label{estimation}

%To estimate the possible number of burst events in the GASTON-GP field, we based our calculations on the number of massive protostars in GASTON-GP field and the current (limited) statistics of accretion bursts from simulations.

Protostellar clusters form within clumps, i.e. the parsec-scale over-dense regions within molecular clouds \citep{Williams2000}. Within clumps, cores host the formation of individual stars and small stellar systems with typical sizes of $0.01-0.1$~pc \citep{Andre2000}. The GASTON-GP compact sources identified here have a mean size of $\sim0.7$~pc and can therefore be classified as clumps. The star formation efficiency in such clumps is believed to be of the order of $\sim10\%$ \citep{Saito2007, Higuchi2009, wells2022}. With a total GASTON-GP 1.15 mm compact source mass of $1.82\times10^{5}$ \sm, we expect these clumps to form a total of $1.82\times10^{4}$ \sm\ worth of stars.
Assuming a Kroupa IMF \citep{2001MNRAS.322..231K}
%: $\frac{dN}{dm} \propto m^{-\alpha}$, where $\alpha$ is 0.3 when $0.01\le m/M_{\odot}<0.08$, 1.8 when $0.08\le m/M_{\odot}<0.5$, 2.7 when $0.5\le m/M_{\odot}<1$ and 2.3 when $1\le m/M_{\odot}$. Considering a 
and a stellar mass range of 0.01 \sm\ to 120 \sm, the identified compact sources are expected to spawn approximately 87 protostars of 10 to 20~\sm , 45 protostars of 20 to 60~\sm\ and 8 protostars of 60 to 120~\sm.

\citet{Meyer2017, Meyer2019, Meyer2021} investigated accretion bursts of massive protostars using hydrodynamic  simulations of core collapse. In \citet{Meyer2021}, the authors present the statistical properties of accretion bursts occurring during massive star formation under different initial conditions, such as different initial core masses and ratios of rotational-to-gravitational energy. Statistics they provide cover the frequency, duration, time interval, total burst time and accreted mass under different magnitude cut-off thresholds (1 to 4 magnitude, i.e., 2.5 to $2.5^4$ times increase in luminosity). Here, we will mostly make use of the burst frequency as a function of the final stellar mass, at a cut-off threshold of different magnitude. 
To estimate the maximum number of bursts, we made the optimistic assumption that all luminosity bursts of 1 magnitude or more are detectable in the GASTON-GP data. 
% This is supported by the GASTON-GP detection of the G24.33+0.14 (S250 in Tab. \ref{candidate_tab}), which has a burst of 1.3 magnitude in NEOWISE W2 band. Though the exact luminosity magnitude of G24.33 + 0.14 is not yet known, the much better correspondence of variability at the infrared band and the corresponding $\sim 15\%$ flux variation at 1.15 mm support the 1 mag threshold. From the modelling in following Sec. \ref{beam_effect}, a luminosity burst of 1.38 to 4 times (under Case 1 and Case 2) can cause a $15\%$ flux variation ta 1.15 mm.
In \citet{Meyer2021}'s simulations, after a total collapse time from $\sim$ 30 to 60 kyr, stars with masses ranging from $10 M_{\odot}$ to $60 M_{\odot}$ formed. On average, a $10 M_{\odot}$ protostar experiences 12 bursts for a total accretion burst time of 238 years during 60 kyr in the simulations. A $20 M_{\odot}$ one would experience 46 bursts for a total burst time of 740 years during 59.6 kyr. And a $60 M_{\odot}$ one typically experiences 187 bursts for a cumulated accretion burst time of 735 years during 35.1 kyr. Note all the bursts included here are brighter than 1 magnitude in luminosity.

To derive the expected number of bursts within the GASTON-GP field we combine statistics from \citet{Meyer2021}'s simulations to the number of massive stars we expect to be formed (see above). For that purpose, and due to the limited number of stellar masses produced in those simulations, we assumed that 10–20 \sm\ stars follow the burst characteristics of the 10 \sm\ model, 20–60 \sm\ stars follow the 20 \sm\ model, and 60–120 \sm\ sources follow the 60 \sm\ model. The number of individual bursts expected to occur within the GASTON-GP field is given by: 
\begin{equation}
N_{\rm exp} = \sum_{m} n_{{\rm proto}, m}\times n_{{\rm burst},m}\times \Delta{t} / T_{m},
\end{equation}
where $\Delta{t}$ is the 4-year observation time, $T_{m}$ is the collapse timescale from the simulations for a $m$ \sm\ protostar, $n_{{\rm proto}, m}$ is the number of protostars going to form at mass range $m$ and $n_{{\rm burst}, m}$ is the average number of bursts in $m$ \sm\ simulations. For the $m=[10, 20, 60]$ \sm\ models, the corresponding values are: $n_{{\rm proto}, m} = [87,\ 45,\ 8]$, $n_{{\rm burst}, m} = [12,\ 46,\ 187]$, $T_{m} = [60.0,\ 59.6,\ 35.1]\ {\rm kyr}$.  That leads to $N_{\rm exp}= 0.38$.  If we now consider the mean duration time of each burst (19.8 yr for bursts of 10 \sm\ protostars, 16.1 yr for 20 \sm\ protostars, 3.9 yr for 60 \sm\ protostars, estimated from \citet{Meyer2021}) and add it to our 4-yr observation window, bursts occurring between $4-t$ and $4+t$ might be observed, where $t$ is the mean burst duration. As a result, the expected number of burst detection increases to $N_{\rm exp}=2$. 
Thus, based on these calculations, we could have expected to detect at least a couple of bursts within the GASTON-GP dataset. 
The lack of detections could be due a burst stochasticity, uncertainty in the burst parameters, or, most likely, due to luminosity burst sensitivity of the GASTON-GP data. Indeed, $N_{\rm exp} =2$ is estimated under the assumption of detecting all luminosity bursts brighter than 1 mag. However, as discussed in Sec. \ref{variable_comparison}, the sources exhibiting $\sim$1.3 mag variability in the NEOWISE W2 band were not detected in the GASTON-GP data. Although factors such as observing cadence, large uncertainties, and smaller flux variations at millimetre wavelengths can contribute to this non-detection, these results imply that the effective number of detectable bursts is most likely to be significantly lower.

\subsection{Flux variation at 1.15 and 2.00 mm} \label{wavelength_dependence}
Changes in the accretion luminosity of a protostar will lead to wavelength-dependent flux variations. To quantify the level of variation at 1.15 and 2.00 mm, we modelled the spectral energy distribution (SED) of protostellar cores as modified blackbodies:
\begin{equation}
S_{\nu} = \Omega (1 - e^{- \tau_{\nu}}) B_{\nu }(T)
\end{equation}
\noindent where $S_{\nu}$ is the flux density at frequency $\nu$, $\tau_{\nu}$ is the optical depth from the source to the observer, $B_{\nu }(T)$ is the Planck function at temperature $T$, and $\Omega$ is the solid angle subtended by the source. In the millimetre regime, a modified blackbody model is an accurate description of the dust emission. For blackbody radiation, the dust temperature can be estimated from the total luminosity $L_{\rm tot}$ via $T_{\rm dust} \propto L_{\rm tot}^{0.25}$. From this relation we scaled the SED temperature to a given increase in luminosity. For example, a 100-fold increase in luminosity corresponds to a dust temperature that is a factor $\sim3$ larger.

\begin{figure}
\includegraphics[scale=0.7]{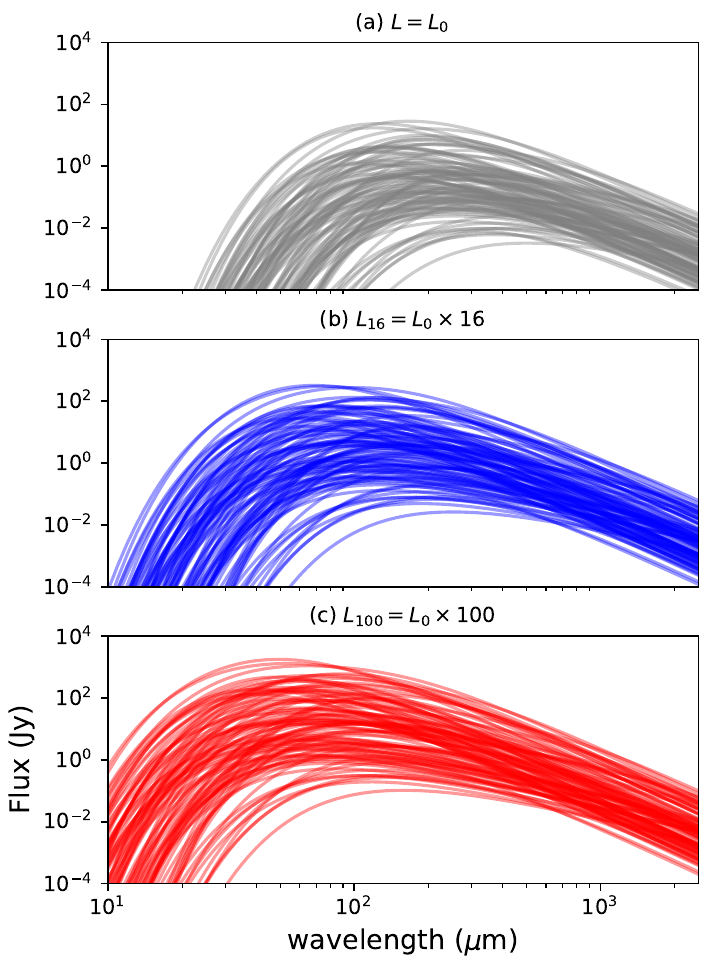}
\caption{{\bf Panel (a):} SED of 100 modelled cores. {\bf Panel (b):} SED for the same set of cores whose luminosities (via their temperatures) increased by a factor of 16. {\bf Panel (c):} SED for the same set of cores whose luminosities (via their temperatures) increased by a factor of 100.}
\label{SED_example}
\end{figure}

We modelled the SEDs of a set of 100 cores with parameters sampled from an uniform distribution. The temperatures range from 10 to 30 K, radii are from 0.001 to 0.05 pc, and masses are from 1 to \(250\ M_{\odot}\), respectively. These parameter ranges were chosen based on the distribution of core properties in \citet{Motte2022} and \citet{Rigby2021a}. Their distances were set to 5 kpc. In the top panel of Fig.~\ref{SED_example}, we present the SEDs of the 100 cores with original luminosity $L_0$. In the middle and bottom panel, we present the SEDs for the same set of cores after increasing the luminosities to 16 and 100 times, respectively. 

\begin{figure}
\includegraphics[scale=0.7]{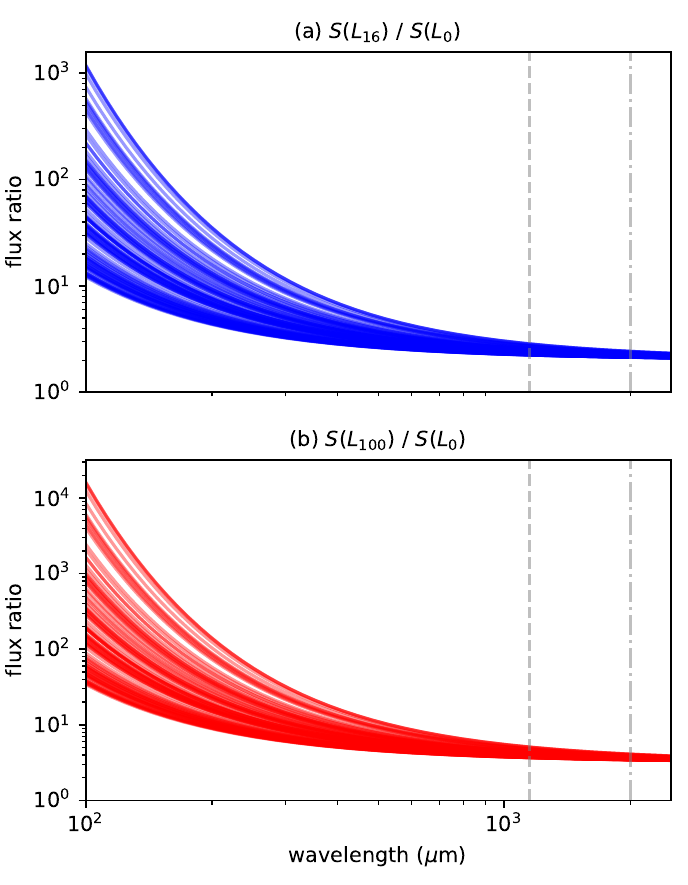}
\caption{{\bf Panel (a):} Flux ratio when the luminosity increased by a factor of 16. {\bf Panel (b)} Flux ratio when the luminosity is increased by a factor of 100. The dashed line and dash-dotted line marks 1.15 mm and 2.00 mm, respectively.}
\label{SED_ratio}
\end{figure}

In Fig.~\ref{SED_ratio}, we compared the SEDs and presented the flux ratio \(S_{\lambda}/ S_{\lambda,0}\) at wavelength $\lambda$. 
Both panels show significant decreases in the flux ratio as the wavelength increases, indicating significantly smaller flux variations at longer wavelengths. When the luminosity increases by a factor of 16 (top panel), the flux ratio lies between 2.2 and 2.85 at 1.15 mm, and between 2.1 and 2.4 at 2.00 mm. When the luminosity is increased by a factor of 100 (bottom panel), the flux ratio varies between 3.6 and 5.1 at 1.15 mm and from 3.4 to 4.1 at 2.00 mm. While the magnitude of the millimetre flux variation is an order of magnitude lower than the corresponding luminosity change, a factor of 2 of flux increase is significantly larger than both the absolute  (14\% and 7\% at 1.15 mm and 2.00 mm) and relative (3.4\% at 1.15 mm and 2.6\% at 2.00 mm) calibration uncertainty of the GASTON-GP data. In fact, a 20\% increase of the 1.15 mm flux, i.e. a $5\sigma$ to $6\sigma$ variability detection, would correspond to a factor of $\sim2$  increase in luminosity. This implies that even relatively faint, {\it isolated}, luminosity bursts could, in principle, be detectable in the GASTON-GP observations. Still, we have not detected any. By comparing the flux ratios at different wavelengths in Fig. \ref{SED_ratio}, we find that the variability amplitudes are significantly larger at shorter wavelengths. This is consistent with observational results, where most detected accretion outbursts and variability in high- and intermediate-mass star-forming regions are reported at infrared wavelengths, reflecting the stronger response of shorter wavelengths to accretion-driven luminosity changes \citep{Caratti2017, Hunter2017, Chen2021, Stecklum2021, Hirota2022, Wolf2024, Park2024, Chen2025}.

\subsection{Beam effects} \label{beam_effect}
The change of a core's luminosity due to accretion occurs at the surface of the protostar's photosphere, i.e. within the inner AU of the core. Such a change, though, will warm up the surrounding core up to a few thousand AU scale \citep{Johnstone2013}, i.e., the typical core-size in massive star-forming regions \citep{2022A&A...662A...8M}. A 2000-AU core sitting at the mean distance of the GASTON-GP sources (i.e. 5.2 kpc) would have an angular size of 0.38$''$, which is much smaller than the GASTON-GP angular resolution of 11.6$''$ and 18$''$ at 1.15 mm and 2.00 mm, respectively. Therefore, within the GASTON-GP beam, there will probably be many unresolved cores. When one of those cores undergoes a luminosity burst, the observed flux variation will be determined by the ratio between the flux variation and the total flux within the beam. If many sources are present, the flux variation and the detectability of the burst will be significantly affected.

At this stage, we do not know the fragmentation level of the GASTON-GP compact sources, but it is most likely that it will be similar to other massive clumps that have been observed with ALMA at high-angular resolution \citep{Csengeri2017, Svoboda2019, Morii2024, Motte2025}. In \citet{Svoboda2019}, they studied a sample of twelve $70~\mu$m-dark high-mass clumps and found them to fragment into 1 to 11 cores at 3000 AU resolution.
We therefore designed three cases to attempt to represent the range of scenarios we might encounter in terms of the fragmentation of GASTON-GP sources.

\noindent {\bf Case 1:} A GASTON-GP compact source is not sub-fragmented and therefore only hosts a single core with a flux $S_0$ (single-core condition);\\ 
\noindent {\bf Case 2:} A GASTON-GP compact source sub-fragments into 5 cores, all with the same pre-burst flux $S_0$. One of the cores is variable; \\
\noindent {\bf Case 3:} A GASTON-GP compact source sub-fragments into 20 cores, all with the same pre-burst flux $S_0$. One of the cores is variable.\\

\begin{figure*}
\includegraphics[scale=0.55]{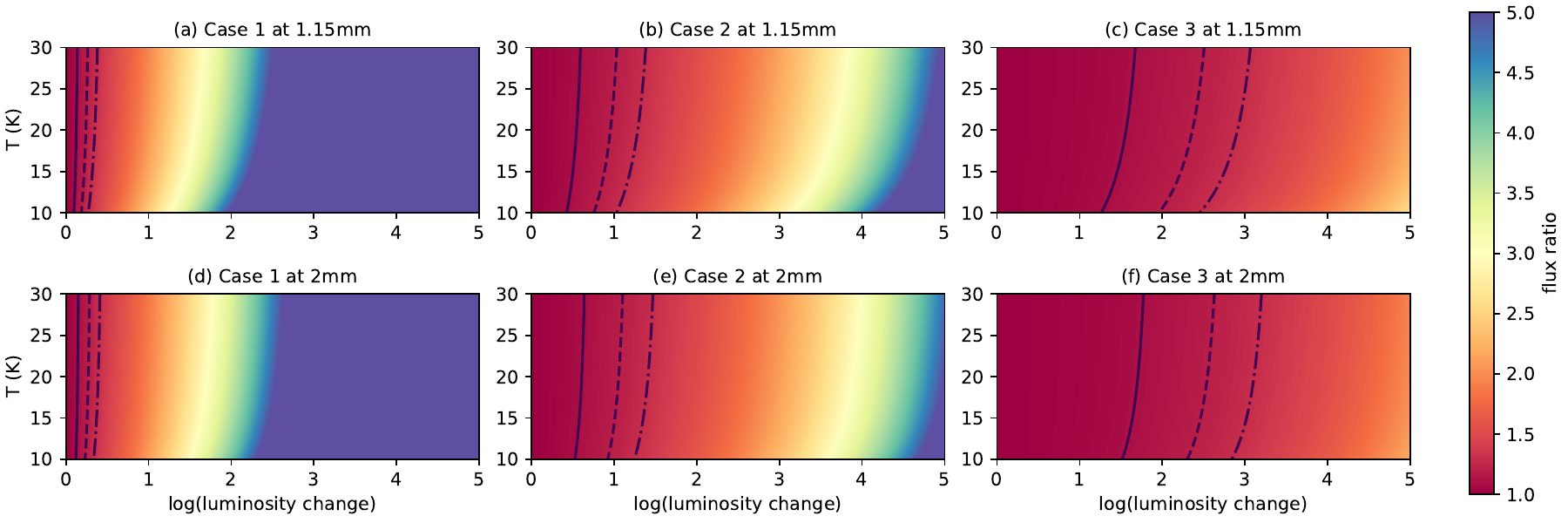}
\caption{ Observable flux ratio under different fragmentation scenarios. {\bf Panel (a):} the observable flux variation at 1.15 mm in Case 1. The image demonstrates the observable flux variation under different temperature $T$ and different luminosity variation. The solid, dashed and dash-dotted line marks the $10\%$, $20\%$ and $30\%$ observable flux variation, respectively. {\bf Panel (b):} the observable flux variation at 1.15 mm in Case 2. {\bf Panel (c):} the observable flux variation at 1.15 mm in Case 3. {\bf Panel (d):} the observable flux variation at 2.00 mm in Case 1. {\bf Panel (e):} the observable flux variation at 2.00 mm in Case 2. {\bf Panel (f):} the observable flux variation at 2.00 mm in Case 3. The description of three cases can be found in Sec. \ref{beam_effect} and the statistics can be found in Tab. \ref{115table}.}
\label{flux_ratio}
\end{figure*}

For each case, we first constructed a parameter space defined by temperature and luminosity variations, with temperature ranging from 10 to 30 K and luminosity variation ranging from 1 to $10^5$. For each grid point within this space, we calculate the pre-burst and post-burst flux densities, from which we derive the flux ratio at both 1.15 mm and 2.00 mm. For Case 1, we place this variable core on its own within a GASTON-GP beam. For Case 2 and 3, we add 4 and 19 non-variable sources with the same pre-burst flux density as the variable one.
The results of this simple calculations are plotted in Fig. \ref{flux_ratio}, with detection statistics presented in Tab. \ref{115table}. As expected, Case 1 is the most favourable one whereby a 20\% flux variation at 1.15 mm will be caused by a luminosity increased by a factor of 1.5. However, if we want to detect the same level of flux variation in Case 2 and Case 3 with more cores within the beam, the luminosity needs to increase by a factor of 6 and 91, respectively. At 2.00 mm, the factors of luminosity change are 1.7, 8, and 200 to produce a 20\% flux variation in Case 1, 2, and 3. The lack of detection of variable protostellar sources in the GASTON-GP field thus indicates the absence of $\sim$ 100-fold, or higher, luminosity bursts between 2017 and 2021. Such bursts are rare, simulations indicating that on average, only 5 bursts larger than 4 mag typically occur during the collapse of a massive core \citep{Meyer2019}. If we were only able to detect such bright bursts, then the expected number of detections during the GASTON-GP observations would become $N_{\rm exp}=0.15$ after we consider the burst duration for 20 and 60 \sm models (see Sec. \ref{estimation}), becoming then fully consistent with our findings.

These models clearly indicate that detecting accretion bursts towards Galactic plane sources at millimetre wavelengths with single-dish continuum data is challenging. Repetitive, sensitive and high-resolution observation of individual cores ($\sim0.5''$ for regions located at $\sim5$~kpc) are needed to significantly increase the number of burst detections. With the increased continuum sensitivity of upcoming single-dish facilities like AtLAST and interferometers such as NOEMA and ALMA, such observations will be possible \citep{Klaassen2020}.  
%Our study clearly demonstrates the large potential of those repetitive continuum Galactic plane surveys that might come to light with AtLAST \citep{Klaassen2019}, which has a mapping speed hundreds of times greater than current large-aperture telescopes \citep{Klaassen2020}. Its higher resolution (< 5$''$ at 1.1 mm) and sub-mm operation wavelength may also largely increase the possibility of detecting accretion bursts. 
%\textcolor{red}{[do we need it]Thanks to its fast mapping speed, an AtLAST survey of Galactic plane (sub)millimetre variables could potentially identify hundreds of protostellar accretion bursts within the same observing time as GASTON-GP.}

\begin{table*}
\label{Lafter-table}
\caption{{\bf Detectability threshold for the three cases under different fragmentation.} The Luminosity values illustrate the threshold for a certain variation detection under different wavelengths and cases from Fig. \ref{flux_ratio}.}
\begin{tabular}{|c|c|c|c|c|c|c|}
\hline
Case                                                           & 10\% {\rm rise in}  $S_{1.15}$ & 20\% {\rm rise in} $S_{1.15}$ & 30\% {\rm rise in} $S_{1.15}$ & 10\% {\rm rise in} $S_{2}$ & 20\% {\rm rise in} $S_{2}$ & 30\% {\rm rise in} $S_{2}$ \\ \hline
\begin{tabular}[c]{@{}c@{}}Case 1\\ (single core)\end{tabular}  &       1.2$L_0$      &  1.5$L_0$           &          1.9$L_0$  &          1.3$L_0$ &     1.7$L_0$      &    2.2$L_0$       \\ \hline
\begin{tabular}[c]{@{}c@{}}Case 2\\ (5 cores)\end{tabular}     &  2.7$L_0$           &    6.0$L_0$         &    10.7$L_0$         &   3.3$L_0$        &      8.4$L_0$     &     17.5$L_0$      \\ \hline
\begin{tabular}[c]{@{}c@{}}Case 3\\ (20 cores)\end{tabular}    &   18.2$L_0$         &  91.4$L_0$           &    285.0$L_0$         &     32.7$L_0$      &    200.0$L_0$       &    691.8$L_0$       \\ \hline
\end{tabular}
\end{table*}

%% file: tex/6_Summary.tex
\section{Summary}

In this paper, we present the complete data release and key results from the GASTON-GP project, which provides deep 1.15 and 2.00 mm continuum maps of a ~2.4 deg$^2$ region in the Galactic plane. The field was observed over 11 runs between 2017 and 2021 with the IRAM 30-m telescope. We combined those observations to construct catalogues and derive the properties of GASTON-GP dense clumps. We then used all individual runs to perform a variability study among $\sim$ 200 high-SNR sources. Our main findings are summarised as follows:

\begin{itemize}
    \item We compiled two compact source catalogues at both 1.15 and 2.00 mm, with 2925 and 1713 sources, respectively. Using NH$_3$ and several CO isotopologues, we determined radial velocities for all sources. Other key physical properties such as the kinematic distance, radii, dust temperature, and mass at different wavelengths were derived. These clumps have a median distance of 5.2 kpc and a median mass of 20 \sm, highlighting the sensitivity of the GASTON-GP observations.

    \item We designed and applied a dedicated relative calibration scheme to the GASTON-GP data to reach a 3.4\% and 2.6\% uncertainty at 1.15 mm and 2.00 mm, respectively. We constructed the light curves for $\sim 200$ high-SNR sources. Under the criteria of having significant variation at least at one wavelength and having correlated light curves at both wavelengths, no robust variable candidates were identified.

    \item We identified a 2.00 mm variable source at a level of 6.4$\sigma$. This source is weaker at 1.15 mm and brighter ($\sim100$ mJy) at 1.3~GHz. It is also isolated from any dusty clumps. Combined, this clearly suggests that this source is not protostellar. Follow-up observations will be needed to characterise its nature.

   % \item \textcolor{red}{We checked the GASTON-GP light curves for three known variable sources with infrared brightening or Methanol maser flares. Unfortunately, no variable signals were found, which can be explained by the large calibration uncertainty, limited cadence or weaker variability at millimetre.}

    \item Based on the simulated burst statistics from {\it isolated} sources, $\sim2$ bursts (> 1 mag) from are expected to occur in GASTON-GP region during 4 years. 

    \item By modelling the SEDs of {\it clustered} protostellar cores and examining the flux ratio at 1.15 and 2.00 mm, we found that the dilution of the variability signal within the $11''$ beam of the GASTON-GP observations only allows the detection of bright, 100-fold or more luminosity bursts. 

    % \item We cross-matched the variable candidates with infrared catalogues from, e.g., NEOWISE and {\it Herschel}. Among the eight candidates, the confirmed variable source S250 shows similar trends in its millimetre and infrared light curves. However, for the rest of the candidates, no clear similarities are seen.

    % \item We compared the 1.15 mm variable signals in different evolutionary groups based on their infrared near to far infrared point source associations. Although associations between GASTON-GP and NEOWISE sources are uncertain, we found more variable signals among GASTON-GP sources with infrared counterparts, indicating an increase rate of accretion bursts with evolutionary stage. 

    % \item We also find that 28 GASTON-GP 1.15 mm compact sources have 6.7 GHz Class II methanol maser counterparts. The increasing ratio of sources with methanol maser counterparts in groups with infrared emission supports the infrared radiation-pumped mechanism for Class II methanol maser. %Four of the candidates with methanol masers are of high possibility of having accretion bursts.
\end{itemize}

%This study demonstrates the current limitations of ground-based single-dish telescopes in identifying accretion bursts at millimetre wavelengths. It highlights the importance of sensitive, high-cadence, and high-resolution Galactic plane surveys at (sub-)millimetre wavelengths, which are essential for studying the accretion processes and understanding how massive stars acquire their mass.

%% file: tex/table115.tex
% To add the table in the back
\begin{landscape}
\begin{table}

\caption{Catalogue of 1.15 mm GASTON-GP compact sources: examples of 20 out 2925 sources are presented. }
\label{115table}
\begin{tabular}{ccccccccccccccc}
\hline
source\_name     & GGP115\_id   & $I_{\rm peak}$       & $\rm SNR_{peak}$  &  $S_{1.15 {\rm mm}}$      & $d$     & $R$        & $v$        & flag  & $T$       & $M$           & high-SNR   & $s_{\rm var}$   & pvalue & GGP2\_id  \\
                &            & (mJy \ beam$^{-1}$) &          & (mJy)         & (kpc)     & (pc)      & (km s$^{-1}$)   &        & (K)      & (\sm)           &            &       &           &                         \\ \hline

... & ... & ...  & ...  & ...   &      ...     &      ...     &        ...    &   ...     & ... &      ...     & ...   &    ...   &   ...      &      ...                   \\
G24.4700-0.4133 & GGP-1-0180 & 10.10$\pm$1.60  & 6.3      & 9.65$\pm$1.91   &           &           &            &        & 18.9$\pm$1 &           & 0    &       &         &                         \\
G22.8658-0.4083 & GGP-1-0181 & 27.16$\pm$2.29  & 11.8     & 178.39$\pm$7.83 & 4.61$\pm$1.2  & 0.53$\pm$0.14 & 76.8$\pm$1.3   & 1      & 17.9$\pm$1 & 78.9$\pm$41.7 & 0    &       &         & GGP-2-0095              \\
G22.9850-0.4125 & GGP-1-0182 & 54.96$\pm$2.27  & 24.2     & 105.70$\pm$4.25 & 4.65$\pm$0.63 & 0.29$\pm$0.04 & 77.8$\pm$1.3   & 1      & 17.4$\pm$1 & 49.5$\pm$14.1 & 0    &       &         & GGP-2-0088              \\
G23.9692-0.4100 & GGP-1-0183 & 14.07$\pm$1.68  & 8.4      & 26.52$\pm$3.14  & 5.82$\pm$1.09 & 0.36$\pm$0.07 & 101.4$\pm$1.3  & 4      & 19.3$\pm$1 & 16.8$\pm$6.7  & 0    &       &         & GGP-2-0097              \\
G23.0675-0.4125 & GGP-1-0184 & 24.86$\pm$2.11  & 11.8     & 18.16$\pm$2.11  & 4.66$\pm$0.71 & 0.16$\pm$0.02 & 78.1$\pm$1.3   & 1      & 20.1$\pm$1 & 7.0$\pm$2.3   & 0    &       &         &                         \\
G22.8308-0.4075 & GGP-1-0185 & 55.40$\pm$2.56  & 21.6     & 161.80$\pm$5.45 & 4.5$\pm$1.47  & 0.33$\pm$0.11 & 74.6$\pm$1.3   & 3      & 17.9$\pm$1 & 68.3$\pm$45.0 & 0    &       &         & GGP-2-0096              \\
G23.3292-0.4067 & GGP-1-0186 & 14.58$\pm$1.81  & 8.1      & 55.18$\pm$4.56  & 3.96$\pm$1.0  & 0.35$\pm$0.09 & 64.0$\pm$1.3   & 4      & 18.7$\pm$1 & 16.9$\pm$8.7  & 0    &       &         &                         \\
G23.1442-0.4092 & GGP-1-0187 & 30.51$\pm$2.12  & 14.4     & 43.63$\pm$3.67  & 3.61$\pm$0.94 & 0.21$\pm$0.05 & 56.8$\pm$1.3   & 4      & 19.2$\pm$1 & 10.7$\pm$5.7  & 0    &       &         & GGP-2-0100              \\
G24.3900-0.4083 & GGP-1-0188 & 19.23$\pm$1.69  & 11.4     & 80.04$\pm$4.84  & 2.8$\pm$1.11  & 0.27$\pm$0.11 & 42.9$\pm$1.3   & 3      & 18.7$\pm$1 & 12.3$\pm$9.8  & 0    &       &         & GGP-2-0101              \\
G23.3183-0.4050 & GGP-1-0189 & 14.59$\pm$1.72  & 8.5      & 39.79$\pm$3.82  & 4.09$\pm$1.03 & 0.31$\pm$0.08 & 66.6$\pm$1.3   & 3      & 19.2$\pm$1 & 12.6$\pm$6.5  & 0    &       &         &                         \\
G23.0717-0.4067 & GGP-1-0190 & 23.22$\pm$2.09  & 11.1     & 31.84$\pm$2.86  & 4.71$\pm$0.89 & 0.22$\pm$0.04 & 79.1$\pm$1.3   & 1      & 19.8$\pm$1 & 12.8$\pm$5.1  & 0    &       &         &                         \\
G23.0433-0.4050 & GGP-1-0191 & 22.55$\pm$2.28  & 9.9      & 26.80$\pm$3.33  & 4.45$\pm$1.15 & 0.22$\pm$0.06 & 73.8$\pm$1.3   & 3      & 19.9$\pm$1 & 9.5$\pm$5.1   & 0    &       &         &                         \\
G22.9492-0.4017 & GGP-1-0192 & 15.60$\pm$2.28  & 6.8      & 52.99$\pm$5.37  & 4.53$\pm$1.18 & 0.36$\pm$0.09 & 75.3$\pm$1.3   & 4      & 20.0$\pm$1 & 19.4$\pm$10.4 & 0    &       &         & GGP-2-0099              \\
G23.1292-0.4042 & GGP-1-0193 & 21.92$\pm$2.11  & 10.4     & 47.68$\pm$4.04  & 3.62$\pm$0.95 & 0.24$\pm$0.06 & 57.0$\pm$1.3   & 4      & 19.3$\pm$1 & 11.7$\pm$6.3  & 0    &       &         & GGP-2-0103              \\
G23.0258-0.4050 & GGP-1-0194 & 47.56$\pm$2.21  & 21.5     & 83.67$\pm$3.24  & 4.55$\pm$1.17 & 0.22$\pm$0.06 & 75.8$\pm$1.3   & 1      & 16.8$\pm$1 & 39.3$\pm$20.5 & 0    &       &         & GGP-2-0102              \\
G22.8492-0.4008 & GGP-1-0195 & 22.16$\pm$2.37  & 9.4      & 75.33$\pm$5.73  & 4.56$\pm$0.54 & 0.37$\pm$0.04 & 75.8$\pm$0.1 & 0      & 18.5$\pm$1 & 31.2$\pm$8.1  & 0    &       &         & GGP-2-0105              \\
G24.9725-0.4042 & GGP-1-0196 & 15.89$\pm$2.45  & 6.5      & 14.52$\pm$2.85  & 5.8$\pm$2.07  & 0.23$\pm$0.08 & 101.5$\pm$1.3  & 3      & 18.2$\pm$1 & 10.0$\pm$7.4  & 0    &       &         &                         \\
G23.0817-0.4033 & GGP-1-0197 & 53.54$\pm$2.21  & 24.2     & 81.19$\pm$3.40  & 4.68$\pm$0.74 & 0.25$\pm$0.04 & 78.5$\pm$1.3   & 1      & 18.0$\pm$1 & 36.8$\pm$12.1 & 0    &       &         & GGP-2-0104              \\
G22.9100-0.3983 & GGP-1-0198 & 29.40$\pm$2.26  & 13.0     & 108.01$\pm$6.03 & 4.49$\pm$1.17 & 0.41$\pm$0.11 & 74.4$\pm$1.3   & 1      & 18.7$\pm$1 & 42.5$\pm$22.5 & 0    &       &         & GGP-2-0107              \\
G23.0092-0.4000 & GGP-1-0199 & 151.61$\pm$2.27 & 66.8     & 175.60$\pm$2.82 & 4.66$\pm$1.22 & 0.19$\pm$0.05 & 78.1$\pm$0.1 & 0      & 16.6$\pm$1 & 88.5$\pm$47.0 & 1    & 1.6   & 0.82    & GGP-2-0106                \\
... & ... & ...  & ...  & ...   &      ...     &      ...     &        ...    &   ...     & ... &      ...     & ...   &    ...   &   ...      &      ...                   \\ \hline
\end{tabular}
\\
\begin{tablenotes}
\item Notes for each column: \\
{\bf source\_name:} Source name based on the peak-intensity location in Galactic coordinates.\\
{\bf GGP115\_id:} GASTON-GP identification name. GGP stands for GASTON-GP, 1 and 2 stand for 1.15 mm and 2.00 mm, respectively. The last number corresponds to the compact id.\\
{\bf $I_{\rm peak}$ }: Peak intensity on 1.15 mm deep map and its uncertainty.\\
{\bf $\rm SNR_{peak}$}: SNR at the peak location on 1.15 mm deep map.\\
{\bf $S_{1.15 {\rm mm}}$}: Integrated flux density at 1.15 mm and its associated uncertainty, estimated as ${\rm mean(rms)} \times \sqrt{n_{\rm beam}}$, where $n_{\rm beam}$ is the number of beams within the source footprint.\\
{\bf $d$}: Kinematic distance assigned to the source, updated using maser parallaxes.\\
{\bf $R$}: Size of the compact sources. Uncertainties come from the distance uncertainty.\\
{\bf $v$ }: Assigned radial velocity of the source after the branch check (see text and details in App. \ref{velocity_decision}). The velocity uncertainty is 0.1 (0.1 in the table) km s$^{-1}$ for velocities from NH$_3$ and 1.3 km s$^{-1}$ for those from CO. Both the numbers correspond to the observation resolution.\\
{\bf flag}: Flag of the radial velocity. 0 means the velocity is derived from NH$_3$, 1 to 2 mean the velocity is derived from C$^{18}$O, 3 to 5 mean the velocity is derived from $^{13}$CO, and 6 and 7 mean the velocity is derived from $^{12}$CO. \\
{\bf $T$ }:Temperatures of compact sources measured from temperature map from \citep{Peretto2016}. A temperature uncertainty of 1 K is adopted for all the sources.\\
{\bf $M$}: Mass of each compact sources. The uncertainty is estimated by combining the contributions from the flux density, kinematic distance, and temperature.\\
{\bf high-SNR}:This column highlights whether or not a source is one of the high-SNR sources for variability study, where 0 means that it is not, and 1 mean it is. \\
{\bf $s_{\rm var}$}: The maximum variable significance compared to the reference intensity. Only given for high-SNR sources.\\
{\bf pvalue}: The p-value of Pearson correlation between light curves at both wavelengths. Only available when the source is a high-SNR source at both wavelengths.\\
{\bf GGP2\_id}:GASTON-GP identification name at 2.00 mm.

(The complete version of this table is available online.)
\end{tablenotes}
\end{table}
\end{landscape}

%% file: tex/A_transfer_function.tex
\section{Transfer Function}
\label{app:transferfunction}

In Section \ref{data} we described the data reduction procedure used to produce the deep maps using the {\sc piic} software. Here, we describe how to estimate the so-called transfer function, which measures the fraction of flux that is recovered by the data processing system relative to the astronomical coming in to the Earth's atmosphere. The general idea is to produce a model of the total sky emission for both NIKA2 bands, insert the appropriate signals into raw timelines, and then process those time-series data using the same {\sc piic} pipeline as was used to reduce the deep maps. By comparing the output of the pipeline to the input, we should have an idea of what level of spatial filtering is being applied by the system.

We created models of the emission at 1.15 and 2.00\,mm by assuming thermal dust emission following a modified blackbody:
\begin{equation}
I_\nu = \mu_{\mathrm{H}_2} m_\mathrm{H} N_{\mathrm{H}_2} \kappa_\nu B_\nu (T_\mathrm{col})
\end{equation}
\noindent where again we use the colour temperature maps of \citet{Peretto2016}, along with their H$_2$ column density counterparts as inputs for our model. Again, $\kappa_\nu$ is calculated as 0.1 cm$^2$\,g$^{-1}$ $(\nu /\mathrm{1\,THz)}^{\beta}$. Finally, we rescale the resulting models by factors of 0.5 and 0.4 at 1.15 and 2.00\,mm, respectively, because we find that without this rescaling, our models overestimate the sky flux by the corresponding amount. This \emph{Herschel}-based model has an angular resolution of 18 arcseconds, which is larger than that of the 1.15\,mm NIKA2 imaging, and comparable to the resolution of NIKA2 at 2.00\,mm, and so this must be borne in mind in this analysis.

The transfer function estimate in {\sc piic} has two important aspects. First, we use the \verb|nullMap| option, which causes the pipeline to multiply every second scan by $-1$ such that when the timelines are eventually combined, the sources of signal common to both scans (i.e. astronomical source) ought to cancel out, producing a null map (NM) which contains only the appropriate noise field. However, in practice, the null maps have different characteristics to the noise field in the final maps, due to variations in observing conditions (elevation, scanning direction, zenith opacity, calibration). Our null maps were produced by carefully pairing the scans to match as closely as possible, but often they would be from different days. The null maps, therefore, contain residual emission, but we consider them to be close enough such that the null map, when combined with our emission model, give a realistic enough simulation of the real data. Second, we enable the \verb|addSource| option of {\sc piic]}, which allows us to feed in our \emph{Herschel} model directly to the timelines, which are then reduced in the same way as the deep map. The resulting pipeline-reduced model is called the `null map add source' (NMAS) map.

\begin{figure*}
    \centering
    \includegraphics[width=\linewidth]{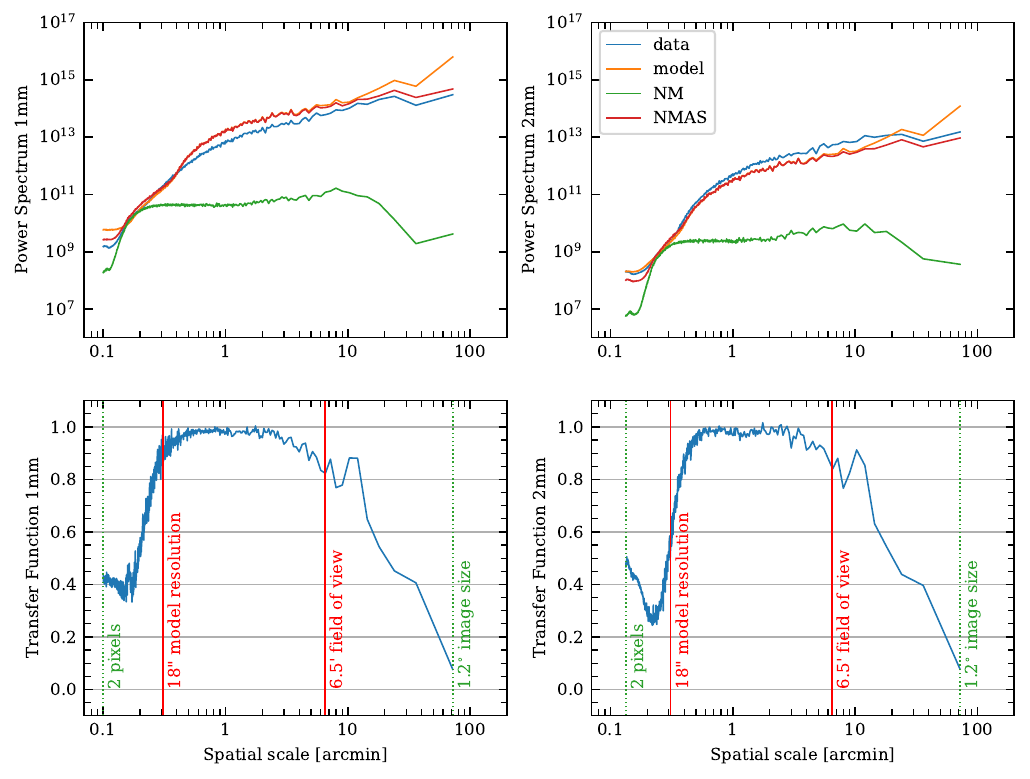}
    \caption{Power spectra (top row) and transfer functions (bottom row) for the 1.15\,mm (left column) and 2.00\,mm (right column) GASTON-GP deep maps, estimated as described in the text.}
    \label{fig:transferfunction}
\end{figure*}

In Fig. \ref{fig:transferfunction} we display the power spectra of the various maps, and the corresponding pipeline transfer functions. The transfer function is measured as 
\begin{equation}
    T(k) = \frac{P_\mathrm{NMAS}(k) - P_\mathrm{NM}(k)}{P_\mathrm{model}(k)},
\end{equation}
where the terms correspond to the power spectra for each of the NMAS, NM, and model images. The power spectra were obtained by measuring the power in the fast Fourier transform in azimuthally-averaged bins, and we have transformed the native spatial frequency axis (which has units of pixel$^{-1}$) to spatial scale in arcminutes to provide the transfer function in a more intuitive format. We highlight various significant scales. The transfer function shows that we accurately recover power between scales of up to $\sim 3$ arcminutes at the 95 per cent level, and at the scale of the 6.5 arcminute field of view, we still recover $\sim 85$ per cent of the power. At scales larger than this, the power is recovered less reliably, though the significant power remaining is, we believe, due to source crowding along the Galactic plane such that there is no significant emission-free background region. It is also worth noting the transfer function drops off at scales smaller than $\sim$40 arcsec. We expect this behaviour below the scale of the model resolution (18 arcseconds), and suspect that the drop-off in the 18--40 arcsec range is due to a combination of interactions between {\sc piic} and the model, and imperfections in the NM. It is thus unlikely to be an effect that is present within the deep maps. Ideally we would use a model with much greater angular resolution than both NIKA2 beams, but we could not concoct such a model without sacrificing a realistic recreation of the expected source morphologies and intensity distributions.

%% file: tex/B_velocity_decision.tex
\section{Source velocity decision tree}
\label{velocity_decision}

\begin{figure*}
\includegraphics[scale=0.27]{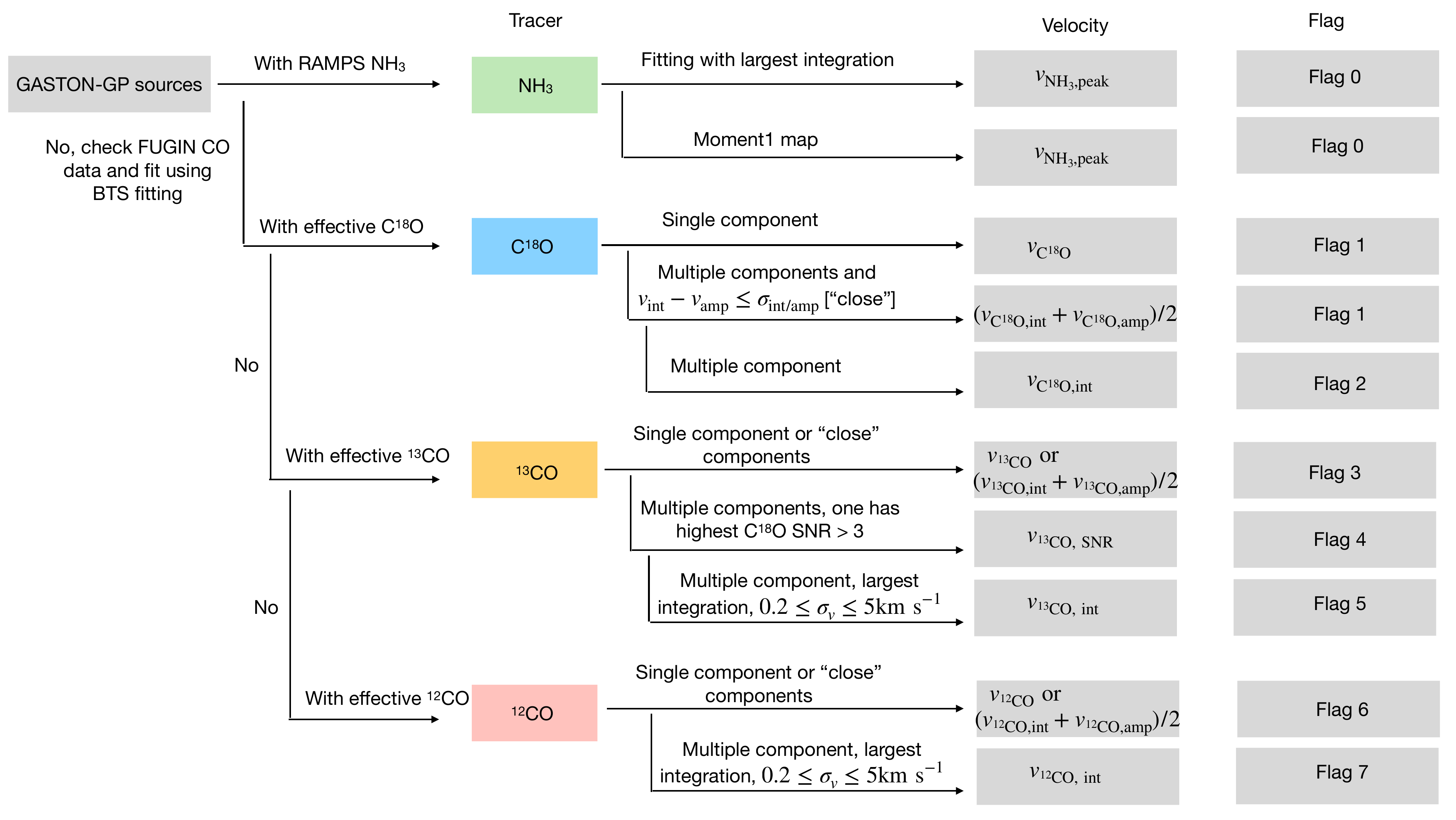}
\caption{Flowchart of the source velocity decision using different tracers.}
\label{velocity_chartflow}
\end{figure*}

To derive the radial velocity of all the compact sources, we used observations of NH$_3$ and the three most common CO isotopologues. Here we detail the procedure used to assign velocities and corresponding flags to each of the leaf structures, and summarise the procedure in Fig. \ref{velocity_chartflow}.

Due to the dense nature of our clumps, we can reasonably expect that those are most accurately identified in the densest gas tracers. We thus first use the NH$_3$ (1, 1) data from the Radio Ammonia Mid-Plane Survey \citep[RAMPS-][]{Hogge2018}. Two types of RAMPS data products are provided in \citet{Hogge2018}: maps of mean velocities arising from fits to the individual spectra (with up to two velocity components per spectrum) and moment 1 maps. We only consider the brightest fitted component in each spectrum and, where available, we take the fitted velocity at the location of each compact source's peak.
For compact sources without fitted velocities from RAMPS, we use the velocity from the NH$_3$ (1, 1) moment 1 map at the peak location. We mark all velocities derived from NH$_3$ with a flag value of 0.

For sources with no RAMPS detection, we use FUGIN \citep{Umemoto2017} C$^{18}$O(1--0), $^{13}$CO(1--0) and $^{12}$CO(1--0) data which fully covers the GASTON-GP field. Considering the possibility of multiple clouds along the line of sight, we first calculated a weighted mean spectrum for each clump within their footprints, where the intensity of the CO emission from each isotopologue in each pixel was weighted by the GASTON-GP intensity. 
The weighted mean spectrum for each source is then given to BTS to fit the velocity components using multiple Gaussian profiles \citep{Clarke2018}. We set {\verb|smoothing_length|} = 3, {\verb|signal_to_noise_ratio|} = 5, {\verb|max_peaks|} = 10, and {\verb|min_velocity_channels|} = 3 in the fitting routing. These parameters specify the channel length used for smoothing, the SNR threshold for fitting, the maximum number of components allowed during fitting, and the minimum number of channels required for a valid fitting, respectively. Due to the fact that $^{12}$CO traces more diffuse gas with a density of $\sim\ 10^2\ {\rm cm^3}$, while $^{13}$CO and C$^{18}$O trace a higher density of $\sim\ 10^3\ {\rm to} 10^4\ {\rm cm}^{-3}$ in the clouds \citep{Scoville1974, Umemoto2017},  we prioritise C$^{18}$O velocities first, then $^{13}$CO and $^{12}$CO last.

For sources with C$^{18}$O(1--0) emission, we examined the fitting results from BTS. If there was a single velocity component, then the corresponding centroid velocity and a flag value of 1 was assigned to the source. Where multiple velocity components were detected, we identified the component with the largest amplitude, and the one with the largest integrated intensity, together with their corresponding centroid velocities $v_{\rm {amp}}$ and $v_{\rm {int}}$, and line widths $\sigma_{\rm amp}$ and $\sigma_{\rm int}$. The mean velocity $(v_{\rm {amp}} - v_{\rm {int}})/2$ would be used if these two components are very close, which can be decided by $|v_{\rm {amp}} - v_{\rm {int}}| \le  \sigma_{\rm amp}$ or $\sigma_{\rm int}$ (i.e. if one component falls within the linewidth of the other), and assigned a flag value of 1.
Otherwise, the mean velocity of the fitting with the largest integral area, $v_{\rm int}$, is chosen and flagged with a value of 2. 

If there is no available fitting from C$^{18}$O(1--0) towards a source, we switched to $^{13}$CO(1--0) and performed a similar analysis on that line, assigning the $v_{\rm ^{13}CO}$ or $(v_{\rm {amp}} - v_{\rm {int}})/2$ and a flag value of 3 for single or very close $^{13}$CO(1--0) components. When there are multiple distinct components, 
we check whether the C$^{18}$O SNR is greater than 3 at the position of any of the components and, if so, we assigned the velocity at which that SNR is the greatest, assigning a flag value of 4.
%we consider the signal-to-noise ratio (SNR) of the C$^{18}$O spectral line within the velocity range of each component if the C$^{18}$O spectra is available. We select the component with the highest C$^{18}$O SNR, provided that it is greater than 3, and assign it a flag of 4. 
When there is no significant C$^{18}$O emission, the central velocity of the component with the largest integrated area is used, but then we also impose that 0.2 \kms $< \sigma_{v} <$ 5 \kms so that both spikes and broad components are rejected. Such cases are assigned a flag value of 5. 

Finally, if there was no $^{13}$CO(1--0) velocity available, we performed the same analysis on the $^{12}{\rm CO}$ data, assigning the $v_{\rm ^{12}CO}$ or $(v_{\rm {amp}} - v_{\rm {int}})/2$ and flag values of 6 for single or close $^{12}$CO(1--0) components. For the remainder, the mean velocity of the component with 0.2 \kms $< \sigma <$ 5 \kms and largest integrated intensity and flag values of 7 were assigned.

